\documentclass[journal=jpclcd,manuscript=article]{achemso}

\usepackage[version=3]{mhchem} 

\usepackage[utf8]{inputenc}
\usepackage{bm}
\usepackage{amsmath}
\usepackage{amssymb}
\usepackage{hyperref}
\usepackage{xcolor}
\usepackage{graphicx}
\usepackage{grffile}

\mciteErrorOnUnknownfalse

\newcommand{\reals}{\mathbb{R}}

\newcommand{\cgind}{I}
\newcommand{\fgind}{i}
\newcommand{\ncg}{N}
\newcommand{\nfg}{n}

\newcommand{\mscgresidual}{\bm{\ell}_{\mathrm{FM}}}
\newcommand{\outermscgresidual}{\mathcal{L}_{\mathrm{FM}}}
\newcommand{\pmf}{W}
\newcommand{\mscgforceagg}{\mathcal{F}}
\newcommand{\cgmap}{\mathcal{M}}
\newcommand{\cgmapmat}{\mathbf{M}}
\newcommand{\pcgmapmat}{\underline{\mathbf{M}}}

\newcommand{\pforcemapmat}{\underline{\mathbf{B}}}
\newcommand{\forcemapmat}{\mathbf{B}}
\newcommand{\constraintmat}{\mathbf{C}}
\newcommand{\trajforcemat}{\mathbf{F}}
\newcommand{\trajforcematreshape}{\mathbf{\underline{F}}}

\newcommand{\aaforce}{\bm{f}}
\newcommand{\outerprod}{\otimes}
\newcommand{\ident}{\mathbf{I}}
\newcommand{\nsteps}{n_t}

\newcommand{\calpha}{{C_\alpha}}

\newcommand{\rcg}{\bm{R}}
\newcommand{\rfg}{\bm{r}}
\newcommand{\fcg}{\bm{\mscgforceagg}}
\newcommand{\ffg}{\mathbf{f}}
\newcommand{\potcg}{U}
\newcommand{\potfg}{V}
\newcommand{\nconstraints}{K}
\newcommand{\constraints}{\sigma}
\newcommand{\paramscg}{\bm{\theta}}
\newcommand{\paramsmap}{\bm{\eta}}
\newcommand{\rRaverage}[1]{
    \left \langle {#1} \right \rangle_{\rfg | \rcg}
}
\newcommand{\raverage}[1]{
    \left \langle {#1} \right \rangle_{\rfg}
}
\newcommand{\Raverage}[1]{
    \left \langle {#1} \right \rangle_{\rcg}
}
\newcommand{\forcemap}{\mathcal{B}}

\newcommand{\angstrom}{\textup{\AA}}

\author{Andreas Krämer}
\affiliation{Department of Mathematics and Computer Science, Freie Universit\"{a}t Berlin, Arnimallee 12, 14195 Berlin, Germany}
\altaffiliation{equal contribution}
\author{Aleksander P. Durumeric}
\affiliation{Department of Mathematics and Computer Science, Freie Universit\"{a}t Berlin, Arnimallee 12, 14195 Berlin, Germany}
\altaffiliation{equal contribution}
\author{Nicholas E. Charron}
\affiliation{Department of Physics and Astronomy, Rice University, Houston, TX, USA}
\alsoaffiliation{Center for Theoretical Biological Physics, Rice University, Houston, USA}
\alsoaffiliation{Department of Physics, Freie Universit\"{a}t Berlin, Arnimallee 12, 14195 Berlin, Germany}
\author{Yaoyi Chen}
\affiliation{Department of Mathematics and Computer Science, Freie Universit\"{a}t Berlin, Arnimallee 12, 14195 Berlin, Germany}
\alsoaffiliation{IMPRS-BAC, Max Planck Institute for Molecular Genetics, 14195 Berlin, Germany}
\author{Cecilia Clementi}
\email{cecilia.clementi@fu-berlin.de}
\affiliation{Department of Physics, Freie Universit\"{a}t Berlin, Arnimallee 12, 14195 Berlin, Germany}
\alsoaffiliation{Center for Theoretical Biological Physics, Rice University, Houston, TX, USA}
\alsoaffiliation{Department of Chemistry, Rice University, Houston, TX, USA}
\author{Frank Noé}
\email{frank.noe@fu-berlin.de}
\affiliation{Microsoft Research AI4Science, Karl-Liebknecht Str. 32, 10178 Berlin, Germany}
\alsoaffiliation{Department of Mathematics and Computer Science, Freie Universit\"{a}t Berlin, Arnimallee 12, 14195 Berlin, Germany}
\affiliation{Department of Physics, Freie Universit\"{a}t Berlin, Arnimallee 12, 14195 Berlin, Germany}
\alsoaffiliation{Department of Chemistry, Rice University, Houston, TX, USA}

\title[noise-reducing]{Statistically optimal force aggregation for coarse-graining molecular dynamics}

\abbreviations{CG, MD, PMF, TICA}
\keywords{coarse-graining, force matching, molecular dynamics}

\begin{document}

\singlespacing

\begin{tocentry}
\includegraphics[width=\linewidth]{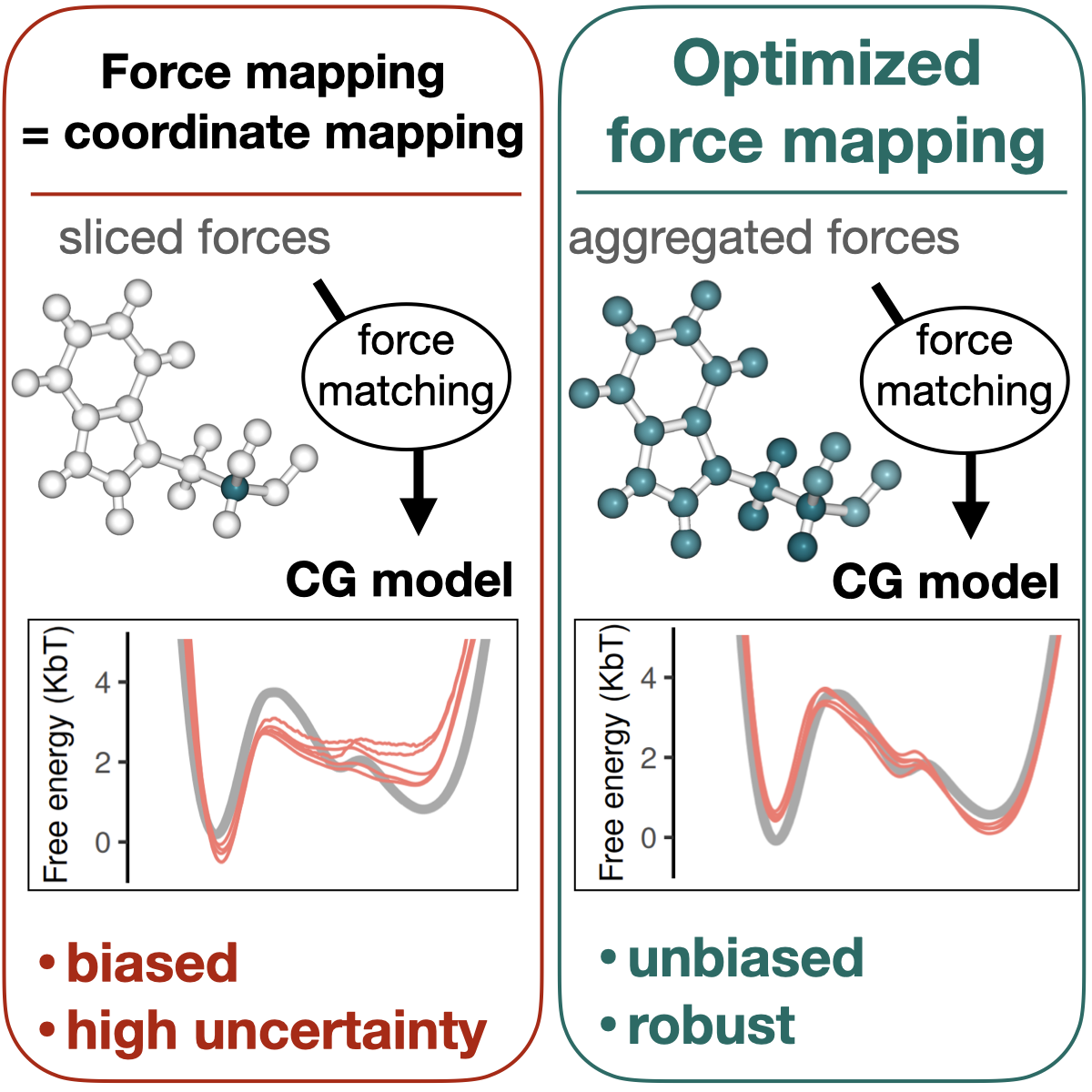}
\end{tocentry}

\begin{abstract}
    Machine-learned coarse-grained (CG) models have the potential for simulating large molecular complexes beyond what is possible with atomistic molecular dynamics. However,  training accurate CG models remains a challenge. A widely used methodology for learning CG force-fields maps forces from all-atom molecular dynamics to the CG representation and matches them with a CG force-field on average. We show that there is flexibility in how to map all-atom forces to the CG representation, and that the most commonly used mapping methods are statistically inefficient and potentially even incorrect in the presence of constraints in the all-atom simulation. We define an optimization statement for force mappings and demonstrate that substantially improved CG force-fields can be learned from the same simulation data when using optimized force maps. The method is demonstrated on the miniproteins Chignolin and Tryptophan Cage and published as open-source code.
\end{abstract}


\section{Introduction}

Atomistic molecular dynamics (MD) simulations provide fundamental insight into physical phenomena by elucidating the behavior of individual atoms.\cite{hollingsworth2018molecular,bottaro2018biophysical,gartner2019modeling} While current simulations scale to millions of atoms and millisecond timescales, their application is constrained by an extremely large computational cost. One leading approach to investigate even larger systems for longer time periods is reducing the computational burden via coarse-graining, where molecular systems are simulated using fewer degrees of freedom than those associated with the atomistic positions and momenta.
Particulate coarse-grained (CG) models typically define CG degrees of freedom (referred to as a beads) as instantaneous averages of multiple atoms\cite{baschnagel2000bridging,klein2008large,noid2013perspective,pak2018advances,dhamankar2021chemically,jin2022bottom}. Once the resolution (i.e., the definition of the CG degrees of freedom) is chosen, the central challenge is finding a force-field that accurately represents the physical interactions that can be used to simulate the complex behavior of large molecular systems. 

Bottom-up coarse-graining focuses on CG force-fields which systematically approximate the CG behavior implied by a reference atomistic force-field\cite{noid2013perspective,dhamankar2021chemically,jin2022bottom}, and has been recently used to parameterize machine-learned CG force-fields based on deep neural networks \cite{lemke2017neural,zhang2018deepcg,Wang2019,wang2019cgautoencoders,Husic2020,wang2021multibody,Chen2021implicit,Chennakesavalu2022ensuring,majewski2022machine,Ding2022coarsegrained,durumeric2023machine,yao2023machine}. However, these applications require large amounts of MD data from the reference atomistic force-field and, in the case of miniproteins, have often not quantitatively reproduced free energy surfaces of high-dimensional reference systems.\cite{lemke2017neural,Wang2019,Husic2020,wang2021multibody,majewski2022machine} These inaccuracies are often attributed to limited data, as the functional forms underpinning the force-field are highly flexible.

\begin{figure}[b!]
    \centering
    \includegraphics[width=0.9\linewidth]{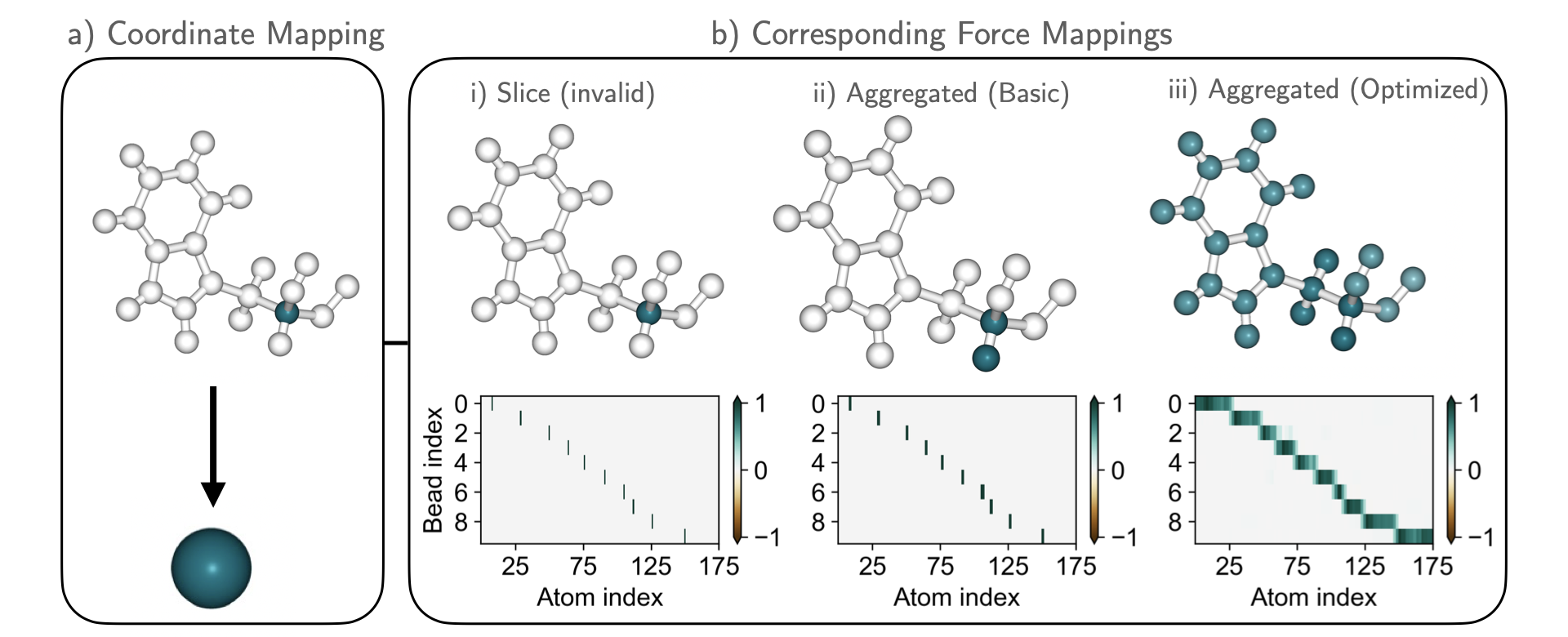}
    \caption{Different force mappings for the same ``slice''  coordinate mapping. This example shows the TRP9 residue of the Chignolin miniprotein. Contributions to the CG bead are color-coded. a) Applying the same slice mapping to forces is invalid due to a rigid bond between the $\calpha$ and  connected hydrogen, and leads to a inaccurate CG force-field. b) ``Basic'' aggregated force mapping, in which the forces of all holonomically constrained atoms contribute with equal weight to the mapped force. c) Statistically optimal force mapping in which all atoms can contribute with weights that are optimized to reduce the statistical uncertainty of the CG force.}
    \label{fig:forcemappings}
\end{figure}

There are multiple approaches to parameterizing bottom-up CG force-fields\cite{noid2013perspective,joshi2021review,dhamankar2021chemically,jin2022bottom}
Unfortunately, many\cite{schommers1973pair,lyubartsev1995calculation,muller2002coarse,toth2007interactions,shell2008relative,cho2009inversion,lu2013fitting,rudzinski2014investigation,schoberl2017predictive,thaler2021learning} of these approaches require the repeated converged simulation of candidate CG force-fields, creating a significant computational barrier to their application in complex systems\cite{Thaler2022deep,durumeric2023machine}. A leading approach  circumventing repeated simulation is Multiscale Coarse-Graining (i.e., ``variational force matching''), where CG potentials are parameterized to directly approximate the effective \emph{mean force} of an atomistic force-field projected to the CG resolution\cite{izvekov2005multiscale,Noid2008,lu2012multiscale}. \citeauthor{Noid2008}\cite{Noid2008} showed that minimizing the mean-squared deviation between a CG candidate force-field and suitably mapped atomistic forces yields the many-body potential of mean force (PMF) and in doing so reproduces to the reference configurational distribution at the appropriate resolution. 

Numerous aspects of the coarse-graining procedure have been studied in depth; we refer readers to recent reviews for a comprehensive overview.\cite{joshi2021review,dhamankar2021chemically,jin2022bottom} For example, work has extensively studied the 
influence of the atom-to-bead mapping \cite{zhang2008systematic,rudzinski2014investigation,Cao2015mscgXI,Foley2015resolution,madsen2017highly,diggins2018optimal,wang2019cgautoencoders,webb2019graphbased,giulini2020informationtheorybased,souza2021martini3,kidder2021energetic,yang2023slicing}, 
functional form of candidate potential \cite{larini2010multiscale,sanyal2016coarse,john2017many,scherer2018understanding,zhang2018deepcg,Wang2019,Husic2020,wang2021multibody,delyser2022coarse}, 
and other details of the fitting routine \cite{dama2013theory,sharp2019multiconfigurational,rudzinski2020coarse,thaler2021learning,jin2021new,Ding2022coarsegrained,Thaler2022deep,sahrmann2022utilizing}. 
However, to our knowledge no work has directly and systematically investigated the influence of the mapping that projects fine-grained (FG) forces to the CG resolution. When considering the theoretical optimization statement defining force matching in the infinite-sample limit, this force mapping only affects a seemingly inconsequential constant offset to the variational statement determining the optimal force-field\cite{Noid2008,jin2022bottom}. However, when learning force-fields in practice, phase space averages are replaced by statistics calculated from MD trajectories to create tractable sample-based variational statements. When the force-field being parameterized is not highly flexible, the distinction between phase space averages and trajectory statistics is often not important. In contrast, when using highly-flexible modern machine-learned force-field representations (e.g. neural networks) this distinction is critical. Parameterizing a machine-learned force-field on a finite trajectory may lead to overfitting: A force-field with optimal performance on a said trajectory may perform poorly on new configurations\cite{mohri2018foundations,Wang2019,durumeric2023machine}. More flexible potentials require more data for their optimization; with a fixed reference trajectory, this imposes an effective upper bound on the complexity of feasible force-fields, limiting application of flexible functional forms.

While difficulties with finite reference data are similarly exhibited with atomistic machine-learned force-field development, the training data used when force matching at the CG resolution contains less information that its atomistic counterpart: energies are not available and forces are noisy\cite{durumeric2023machine}. The noise present in the forces may be an order of magnitude greater than the signal and can be viewed as a major factor in the high data requirements of machine-learned CG force-fields. The present work shows that designing the force mapping to reduce this noise improves trained CG force-fields considerably. We leverage Ciccotti et al. \cite{Ciccotti2005}, showing that the mean force can be obtained via multiple force mappings as long as they obey consistency requirements related to the configuration mapping and molecular constraints in the reference system (Fig.~\ref{fig:forcemappings}). We formulate a variational statement that minimizes the noise of the mapped forces,  significantly improving the signal-to-noise ratio of the force matching training objective. We also show that both high noise and constraint-inconsistent force mappings significantly degrade learned CG force-fields. While these results apply to all force matched CG models, they are especially important for neural network CG potentials, which are sensitive to noise.\cite{flowmatching2023}
An open-source implementation of the proposed force mapping optimization is provided at \url{https://github.com/noegroup/aggforce}.

\section{Theory}

\subsection{Force matching with constraints}
Consider an atomistic system with atom positions $\rfg \in \reals^{3\nfg}$ and a potential energy function $\potfg(\rfg)$ in the canonical ensemble at temperature $T.$ Atomistic holonomic constraints (e.g., rigid bond lengths) are incorporated as a system of equations,
$
        \sigma(\rfg) = \bm{0}.
$

We consider a linear mapping operator 
$
        \cgmap:\reals^{3\nfg} \to \reals^{3\ncg}, \rfg \mapsto \rcg
$
that maps from fine to coarse configurational degrees of freedom. Under mild constraints\cite{Ciccotti2005,Noid2008}, this mapping induces the many-body PMF $\pmf: \reals^{3\ncg} \to \reals$ through the principle of thermodynamic consistency:
    \begin{equation}
        \label{eq:pmf}
        e^{-\beta \pmf(\rcg)}
            \propto  \int e^{-\beta \potfg(\rfg)} 
                \delta (\rcg - \cgmap (\rfg)) 
                \delta (\constraints(\rfg)) 
                d\rfg
    \end{equation}
where $\beta = (k_B T)^{-1}$ and $k_B$ is the Boltzmann constant. 
The integral in Eq. \eqref{eq:pmf} represents a Boltzmann-weighted average over all FG configurations that correspond to a given CG configuration and obey the constraints. Computing this integral over FG states directly is not feasible for most systems of practical interest. Instead, $\pmf$ can be approximated by optimizing over candidate potentials $\potcg(\rcg; \paramscg)$ with tunable parameters $\paramscg$ using variational principles such as relative entropy minimization \cite{shell2008relative} or force matching \cite{Noid2008}.

In force matching, FG positions $\rfg$ and forces $\ffg=-\nabla \potfg(\rfg)$ are recorded from an equilibrium simulation and mapped to the CG space to yield a training dataset of instantaneous force-coordinate pairs $\{(\rcg, \fcg)\}$.
The optimization statement underlying force matching is found by minimizing the mean-squared deviation between model and training forces,
\begin{equation}
    \begin{aligned}
        \label{eq:fm-residual}
        \outermscgresidual (\paramscg) 
        = \raverage{\mscgresidual (\rfg; \paramscg) }
        = \raverage{
        \left\|
             - \nabla_{\rcg} \potcg(\rcg; \paramscg)
            - \fcg
        \right\|_2^2},
    \end{aligned}
\end{equation}
where $\raverage{x} := \int x\ p(\rfg)\ \delta(\constraints(\rfg))  d \rfg$ denotes the thermodynamic average over the FG equilibrium distribution $p(\rfg)\propto e^{-\beta \potfg(\rfg)}$. As previously noted, in practice force-fields are produced by minimizing a sample-based approximation to Eq. \eqref{eq:fm-residual} produced using $\{(\rcg, \fcg)\}$, possibly with regularization\cite{liu2008bayesian,lu2010efficient,Wang2019}. Analogous to the configurational map $\cgmap,$ we need to define a force map that projects atomistic forces to the CG space in such a way that the mapped forces $\fcg$ are an unbiased estimator of the mean force
	\begin{equation}
	    \label{eq:valid_map}
		\rRaverage{\fcg} = -\nabla \pmf(\bm{R}),
	\end{equation}
where we use the notation $\rRaverage{x} := \raverage{x\ \delta(\rcg - \cgmap(\rfg))} / \raverage{\delta(\rcg - \cgmap(\rfg))}$ for conditional averages.

\subsection{Defining valid force mapping operators}

Ciccotti et al. \cite{Ciccotti2005} found the relation between the CG mean force, $-\nabla \pmf(\rcg),$  and the atomistic forces, $-\nabla \potfg(\rfg),$ by differentiating through the analytic expression of the many-body PMF in Eq. \eqref{eq:pmf}.  They showed that the (negative) mean force may be expressed as 
\begin{align}
    \label{eq:mean-force}
    \nabla \pmf(\rcg) 
        &= 
        \Big\langle { 
        	\underbrace{
            	\forcemap(\rfg) \cdot \nabla \potfg(\rfg)
            	- k_B T \,
                \mathrm{div}\,\forcemap(\rfg)
            }_{=-\fcg(\rfg)}
        }\Big \rangle_{\rfg | \rcg},
\end{align}
where $\mathrm{div}\,\forcemap(\rfg) = (\nabla \cdot B_1, \dots,\nabla \cdot B_{3\ncg})^T$ denotes the divergence per CG coordinate and $\nabla$ the Jacobian. The (local) mapping $\forcemap(\rfg) \in \reals^{3 \ncg \times 3 \nfg}$ is a valid force projection if it obeys the following relations
\begin{enumerate}
    \item[(i)]  \emph{Orthogonality to the constraints:}
    \begin{equation}
        \label{eq:cond1_orthogonal}
        \forcemap(\rfg) \cdot \nabla \constraints(\rfg)^T
        = \bm{0}
        .
    \end{equation}
    
    \item[(ii)] \emph{Compatibility with the configurational mapping:}
    \begin{equation}
        \label{eq:cond2_basis}
        \forcemap(\rfg) \cdot \nabla \cgmap(\rfg)^T
        = \ident.
    \end{equation}
\end{enumerate}
Condition $(i)$ ensures that the mapped forces do not act against any atomistic constraints. This is important because rigid constraints do not transmit force information. Thus, the mapping operator  $\forcemap$ must remove spurious (off-manifold) contributions to the force in order to not pollute the mean force computation. Condition $(ii)$ ensures that the force mapping is consistent with many-body PMF induced by the configurational map.

Importantly, Eqs. \eqref{eq:cond1_orthogonal}-\eqref{eq:cond2_basis} define a system of equations for each $\rfg$ that is usually highly underdetermined. This means that the force mapping operator is generally ambiguous for a fixed configurational mapping.  It can even vary as a function of the FG coordinates $\rfg.$ 
Previous work has not made full use of this flexibility. Instead, a common choice to meet condition $(ii)$ is to define the force map as the pseudoinverse\cite{wang2019cgautoencoders,Husic2020,Chennakesavalu2022ensuring}, i.e. $\forcemap = (\nabla\cgmap \cdot \nabla\cgmap^T)^{-1}  \cdot \nabla\cgmap.$
Alternatively, Noid et al. \cite{Noid2008} defined a set of conditions to satisfy both $(i)$ and $(ii)$ in the case of specialized configurational and force mapping operators. They demand that all atoms that are involved in a constraint must contribute with the same force mapping coefficient. Furthermore, atoms must be configurationally uniquely associated to a single bead to have force contributions to that bead. These conditions restrict the design of $\forcemap$ considerably and do not have a solution for some configurational maps when molecular constraints are present (e.g., the slice mappings considered in this article).

To give an example of the actual flexibility of the force mapping operator, consider the setup underlying most of our computational experiments. FG simulations are run with constrained covalent hydrogen bonds as it is typical for biomolecular simulations.\cite{eastman2017openmm} For the configurational mappings we use slice mappings, where bead positions are identical to the positions of selected individual heavy atoms (Fig.~\ref{fig:forcemappings}a). Under the additional conditions that $\forcemap$ not change as a function of configuration and contributions are the same along each spatial component, conditions $(i)$ and $(ii)$ are satisfied by 
    \begin{equation*}
        \pforcemapmat_{\cgind\fgind}
        = \left\{
            \begin{array}{cl}
                 1, & \mathrm{for\ the\ one\ heavy\  atom}\ \fgind\ \mathrm{that\ is\ identified\ with\ bead}\ \cgind, \\
                 0, & \mathrm{for\ heavy\  atoms}\ \fgind\ \mathrm{that\ are\ identified\ with\ a\ different\ bead}, \\
                 \pforcemapmat_{\cgind j}, & \mathrm{for\ all\ hydrogens\ connected\ to\ the\ heavy\ atom\ } j.\\
                 \mathrm{arbitrary} \in \mathbb{R}, & \mathrm{for\ all\ other\ heavy\ atoms}. \\

            \end{array}
        \right.
    \end{equation*}
where we have used $\pforcemapmat_{\cgind\fgind}$ to denote the the static contribution of atom $\fgind$ to CG bead $\cgind$ in $\forcemap$ (see appendix and SI).
The arbitrary coefficients of all heavy atoms which are not identified with or constrained to any CG bead imply considerable flexibility in choosing the force map, which we exploit for noise-reduction.

\subsection{Dual variational principle for force matching and noise-reduction}
As pointed out in previous work,\cite{Wang2019} the force residual in Eq. \eqref{eq:fm-residual} can be decomposed into PMF error and noise. The PMF error represents the bias and variance due to limited expressivity of the CG model and finite data, while the noise represents the inherently stochastic nature of the mapped training forces from the perspective of the CG model. 
When optimizing machine-learned force-fields with force matching, the noise contribution can dominate the force residual, \cite{Wang2019,durumeric2023machine} which leads to high variance and thus data inefficiency and a tendency to overfit.\cite{flowmatching2023}
The inherent flexibility in the choice of force mapping suggests that this situation can be improved by simply switching to a different force mapping scheme. We will therefore search for force maps that both satisfy the consistency relations in Eqs. \eqref{eq:cond1_orthogonal}-\eqref{eq:cond2_basis} and reduce the noise in the gradient estimator associated with the force residual in Eq. \eqref{eq:fm-residual}. To this end, we first derive a new dual variational principle for force matching and noise-reduction. We then use this insight to propose an efficient algorithm to produce forces that make for a more robust training objective.

To formalize the optimization of the force mapping, assume that we have a family of valid force mapping operators $\forcemap(\rfg; \paramsmap)$ that are parameterized by real vector $\paramsmap$.
This means that $\forcemap(\rfg; \paramsmap)$ satisfies conditions $(i)$ and $(ii)$ for all choices of $\paramsmap,$ see SI for such a construction.
Given such a parameterization of force maps, the force matching residual in Eq. \eqref{eq:fm-residual} becomes a function of both the map and the CG potential parameters. 
The integrand of the residual can be decomposed into three components:
\begin{equation}
  \label{eq:decomposition}
  \begin{aligned}
        \mscgresidual (\bm{r}; \paramscg, \paramsmap) 
        &= 
        \left\|
             - \nabla_{\rcg} \potcg(\rcg; \paramscg)
            - \fcg(\rfg; \paramsmap)
        \right\|_2^2
        \\
        &= \mathrm{PMF\ error} (\rcg; \paramscg)
         + \mathrm{noise} (\rfg; \paramsmap)
         + \mathrm{mixed\ term} (\rfg; \paramscg, \paramsmap),
  \end{aligned}
\end{equation}
similarly as in as in \citeauthor{Wang2019}\cite{Wang2019}. While \citeauthor{Wang2019} use these terms to denote averages, we use them here in a pointwise sense, and with a parametrized force map. The mixed term is mean-free (in the limit of infinite sampling)\cite{Wang2019} and the mapped force $\fcg$ is defined as in Eq. \eqref{eq:mean-force}. 
This decomposition is discussed in detail in the SI. Here we summarize the most important implications:
\begin{itemize}
    \item \emph{Consistency of force matching:} The PMF error does not depend on $\paramsmap$. Thus, for any valid force mapping scheme, minimizing the force matching loss $\raverage{\mscgresidual (\bm{r}; \paramscg, \paramsmap)}$ with respect to $\paramscg$ asymptotically yields a many-body PMF (given a sufficiently powerful class of candidate potentials).
    \item \emph{Optimized mapping:} The noise term does not depend on $\paramscg$. Thus, for any guess of candidate potential, minimizing the force matching loss $\raverage{\mscgresidual (\bm{r}; \paramscg, \paramsmap)}$ with respect to $\paramsmap$ gives the same force map.  A perfect, possibly non-linear, zero-noise map would project each atomistic force exactly onto the mean force. 
    \item \emph{Benefit of joint optimization:} The mixed term controls the amount of noise on the parameter gradients. Improving the force map facilitates finding the CG potential and vice versa. 
\end{itemize}
In summary, the symmetry of the generalized force matching residual in Eq. \eqref{eq:decomposition} reflects two orthogonal approaches to approximate the mean force.
The first approach (classic force matching) tries to find the force-field that best explains the atomistic forces. The second approach (noise-reduction) tries to find the mapping that minimizes the variance of the mapped forces. These approaches will 
benefit from each other when used together. In the following section, we exploit this concept by defining force maps that facilitate efficient optimization of the candidate potential.

\subsection{Computationally efficient optimization of linear force mappings}
One way to use this variational principle is the joint optimization of the force residual over $\paramscg$ and $\paramsmap.$
However, such an approach requires significant effort, e.g. computing the expression in Eq. \eqref{eq:mean-force} at each joint optimization step.
Instead, we construct a a configuration independent (``linear'') force map which minimizes the average magnitude of the mapped forces, i.e. we find the optimal map parameters as 
\begin{equation}
    \label{eq:eta_opt}
    \paramsmap_\mathrm{opt} = \underset{\paramsmap}{\arg\min} \raverage{\left\| \fcg(\rfg; \paramsmap) \right\|_2^2}.
\end{equation}
Note that this optimization term has previously been used to select optimal configurational maps\cite{wang2019cgautoencoders}, but not optimal force maps. We algebraically show in the SI that force mapping scheme obtained in this way reduces a bound on the variance of the parameter gradient.
The gradient variance is crucial as neural networks are typically trained using stochastic gradient descent based algorithms, which iteratively follow the parameter gradient estimated on small batches of training examples\cite{montavon2012neural,bottou2018optimization}. The gradient estimated using a single batch can be viewed as a noisy estimate of the gradient that would be obtained by using all the training samples; this noise can slow the training convergence of neural networks.
Significant effort has aimed at reducing the noise generated at each update by utilizing control variates generated from previous optimization iterations\cite{johnson2013svrg,defazio2014saga,schmidt2017sag,nguyen2017sarah,bottou2018optimization}. However, these modified optimization approaches have had limited success when applied to neural networks, likely due to the speed at which optimization iterations diverge from the calculated variates\cite{defazio2019ineffectiveness}.
Eq. \eqref{eq:eta_opt} may be viewed as utilizing control variates in the force averaging procedure to minimize gradient noise; the control variates are the linear combination of various atomistic forces. Unlike existing modifications of stochastic gradient descent, these control variates incorporate information into the training data that would be lost when using a basic, non-optimized force mapping and result in a considerable reduction in variance.

Furthermore, solving Eq. \eqref{eq:eta_opt} is computationally efficient\cite{osqp} and allows us to optimize the mapped forces before optimizing the CG potential. Consequently, the force optimization becomes a part of the data preparation pipeline and we can perform force matching as usual, but with more robust gradients.

\section{Results}
\label{sec:results}

The choice of force mapping can significantly affect the quality of the resulting CG force-field. This is first demonstrated by using a low dimensional CG potential to model a water dimer, which allows us to visualize and discuss the issues caused by atomistic constraints. We then conclude by investigating the effect on high-dimensional CG neural network potentials trained to reproduce the folding behavior of a fast-folding variant of the miniprotein Chignolin  (CLN025) and Trp Cage, systems commonly used to benchmark machine-learned CG force-fields\cite{Wang2019,Husic2020,wang2021multibody,majewski2022machine}.
For both test cases, the supplementary information contains detailed descriptions of the simulations, CG models, and training procedures. 

\begin{figure*}[htbp]
    \centering
    \includegraphics[width=\linewidth]{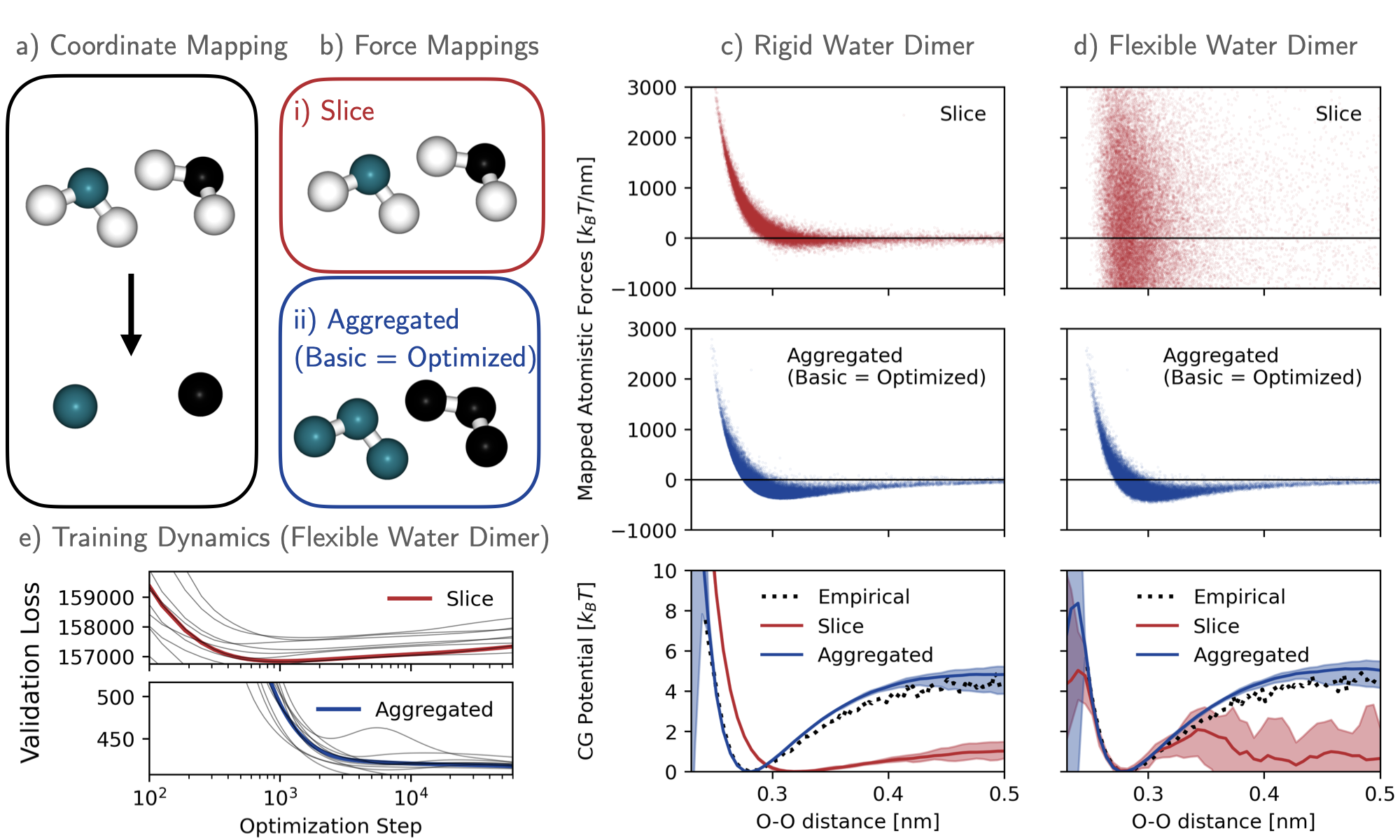}
    \caption{Coarse-graining of water dimers. a) The CG coordinates are defined by retaining only the oxygens. b) Two force mappings were investigated: a slice map and a map with equal weights for oxygens and hydrogens. Turquoise and black represent contributions to bead 1 and 2, respectively. c) Results from constrained atomistic data: forces projected onto the oxygen-oxygen distance through a slice and aggregation mapping. Last row: CG potentials obtained from the projected force data compared to the empirical PMF. d) same as (c) for atomistic data without constraints. e) Mean validation loss during training of the flexible water dimer. The shaded areas in (c-d) represent the values observed over 10 experiments. The grey lines in (e) represent individual experiments.}
    \label{fig:water}
\end{figure*}

\subsection{Water dimers demonstrate the importance of force mappings}

The water dimer system (Fig.~\ref{fig:water}) contains two TIP3P molecules~\cite{jorgensen1983comparison} interacting via Coulomb and Lennard-Jones interactions in a harmonic external potential. Two datasets are created by running MD simulations with and without rigid bond and angle constraints. In both simulations, the most favorable configuration is the dimer state with an oxygen-oxygen distance slightly below 0.3 nm, although distances of up to 3 nm are also explored. 

The configurational mapping and candidate CG force-field basis were fixed: bead positions were identified with oxygen positions (Fig.~\ref{fig:water}a) and the CG potential was defined as a linear combination of radial basis functions on the oxygen-oxygen distance. Two aspects of the coarse-graining task were varied: The force mapping (Fig.~\ref{fig:water}b) and the training data (rigid vs. flexible). We first focus on the rigid system to discuss the influence of atomistic constraints.

\subsubsection{Rigid water: sliced forces are invalid with bond constraints}
Most biomolecular simulations constrain the fastest-moving chemical bonds to enable timesteps greater than $ 1$ fs.\cite{eastman2017openmm} MD engines enforce these constraints by modifying particle positions and velocities at each timestep but do not modify the forces. As a result, the reported forces contain off-manifold contributions, such as spurious radial forces acting along a rigid bond; these artifacts do not influence the atomistic distribution or dynamics but can pollute the force matching objective when not properly taken into account.
The orthogonality condition in Eq. \eqref{eq:cond1_orthogonal} ensures that force mappings eliminate such spurious atomistic contributions to the mapped force. 
The simplest way to enforce this condition is by setting $\pforcemapmat_{\cgind \fgind} =  \pforcemapmat_{\cgind j}$ for any pair of constrained atoms $i$ and $j,$ such that forces felt by atoms connected to atoms preserved in the configurational map via constrained bonds always contribute equally to the mapped force. Throughout this work, we refer to force mappings which only include force contributions from configurationally preserved and their constraint-connected atoms as \emph{basic} (aggregated) force mappings.
For the water dimer with constraints, Fig.~\ref{fig:water}b shows the sliced and basic aggregated force mapping schemes. Slicing in Fig.~\ref{fig:water}b $(i)$ violates the orthogonality condition, while basic aggregation produces valid force mapping for the configurational slice mapping in Fig.~\ref{fig:water}a.

Using invalid force mappings can have a detrimental effect on learning CG force-fields. Fig \ref{fig:water}c shows the mapped forces (using both mapping schemes) versus the bead-to-bead distance. Both force mappings reproduce the intermolecular repulsion at small distances. However, only the basic aggregated forces capture the hydrogen-bond-driven water-water attraction. This flaw is most salient after training CG potentials and evaluating them: potentials trained using basic aggregated forces match the empirical PMF computed from a histogram of the data. In contrast, potentials trained against the sliced forces are inaccurate: they express an overly weak attraction and overestimate the equilibrium distance. This example illustrates how force mappings which violate atomistic constraints can impede convergence to the many-body PMF.

\subsubsection{Flexible water: aggregated forces drive data-efficient coarse-graining}
For the water dimer without constraints, both the slice and basic aggregated force mappings are consistent with the configurational map but they do not both perform equally well. Fig.~\ref{fig:water}d shows forces mapped to the bead-to-bead distance. The sliced forces are dominated by the noise produced by fluctuations in the intramolecular bonds and angles. In contrast, basic aggregation annihilates these contributions completely and greatly reduces the noise in the mapped forces, which is reflected by the magnitude of force matching loss in Fig.~\ref{fig:water}e. Notably, solving the minimization task (Eq. \eqref{eq:eta_opt}) yields the basic aggregation scheme as the optimal linear force mapping (up to a 10$^{-3}$ numerical tolerance). This shows the noise-reduction mechanism at work: aggregating the force over groups of adjacent atoms removes force fluctuations coming from the ``stiff'' local terms of the atomistic potential.

Improving the signal-to-noise ratio of the mapped forces helps train CG potentials on finite datasets. As shown in Fig.~\ref{fig:forcemappings}d, CG models trained on sliced forces only reproduce the mean force in regions where data is abundant, i.e. near the equilibrium distance.
In contrast, models trained on basic aggregated forces yield a high-fidelity approximation to the many-body PMF that agrees well with atomistic statistics.
This result supports the idea that even when slice force mappings are valid given underlying atomistic constraints, using noise-reducing force mappings improves the data-efficiency of creating CG force-fields.

\subsection{Optimized forces improve protein models}

\begin{figure}[htbp]
    \centering
    \includegraphics[scale=0.3]{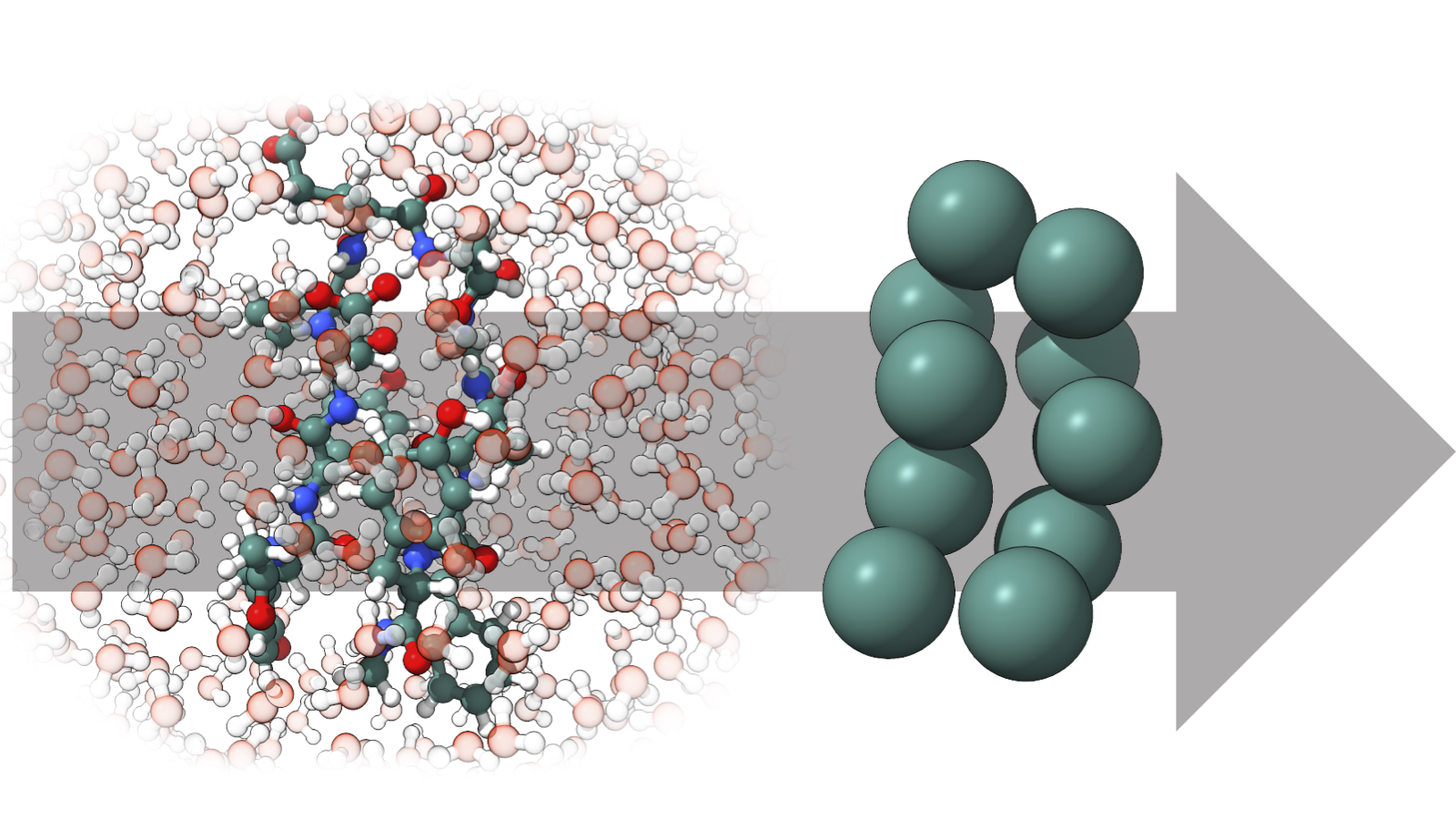}
    \caption{Visualization of the configurational CG mapping used to model Chignolin. The solvated atomistic resolution used for the reference simulations is shown on the left, while the CG representation (which preserves only $\calpha$s) is shown on the right.}
    \label{fig:clncg}
\end{figure}

The proposed force mappings produce significant improvements when coarse-graining proteins use high-dimensional force-fields. Chignolin and Trp Cage, miniproteins consisting of 10 and 20 residues, respectively, exhibit folding behavior and serve as computationally efficient systems for investigating CG force-field design. Here, we model these proteins by only preserving the positions of their $\calpha$ (Fig.~(\ref{fig:clncg})) via the approach described in \citeauthor{Husic2020}\cite{Husic2020} using sliced forces and two modified force mapping operators. 

The reference atomistic simulations utilized constrained bonds to hydrogens; as a result, the sliced force approach, which only includes the forces present on $\calpha$s, is not a valid force mapping for either protein. To investigate valid force mappings we considered two options. First, we tested the basic aggregation force mapping: forces for each CG site were defined as summing the forces of each $\calpha$ with its connected hydrogen(s). Second, we produced an optimized force mapping by solving Eq. \eqref{eq:eta_opt}; this is referred to as the \emph{optimized} mapping (Fig \ref{fig:forcemappings}). Note that water was not considered when creating the optimized force mapping.

The resulting CG force-fields were validated using MD and resulting free-energy surfaces defined along slow coordinates produced via time-lagged independent component analysis (TICA)\cite{Naritomi2011,Perez_JChemPhys2013,Schwantes_JChemTheoryComput2013} on the reference atomistic trajectories. These surfaces were compared to that of the reference atomistic trajectory in three ways. For all approaches, the statistics along the first two TIC components from the model and reference data were histogrammed. In the first approach the difference in the free energy was squared and averaged across bins. For the second approach the Jeffreys divergence (the arithmetic mean of the Kullback–Leibler divergence performed in both directions) was calculated between the two binned distributions. In the third approach, the Jensen-Shannon divergence was similarly calculate between the two binned distributions. Further details on calculating divergences may be found in the SI.

\begin{figure}[htbp]
    \centering\includegraphics[width=\linewidth]{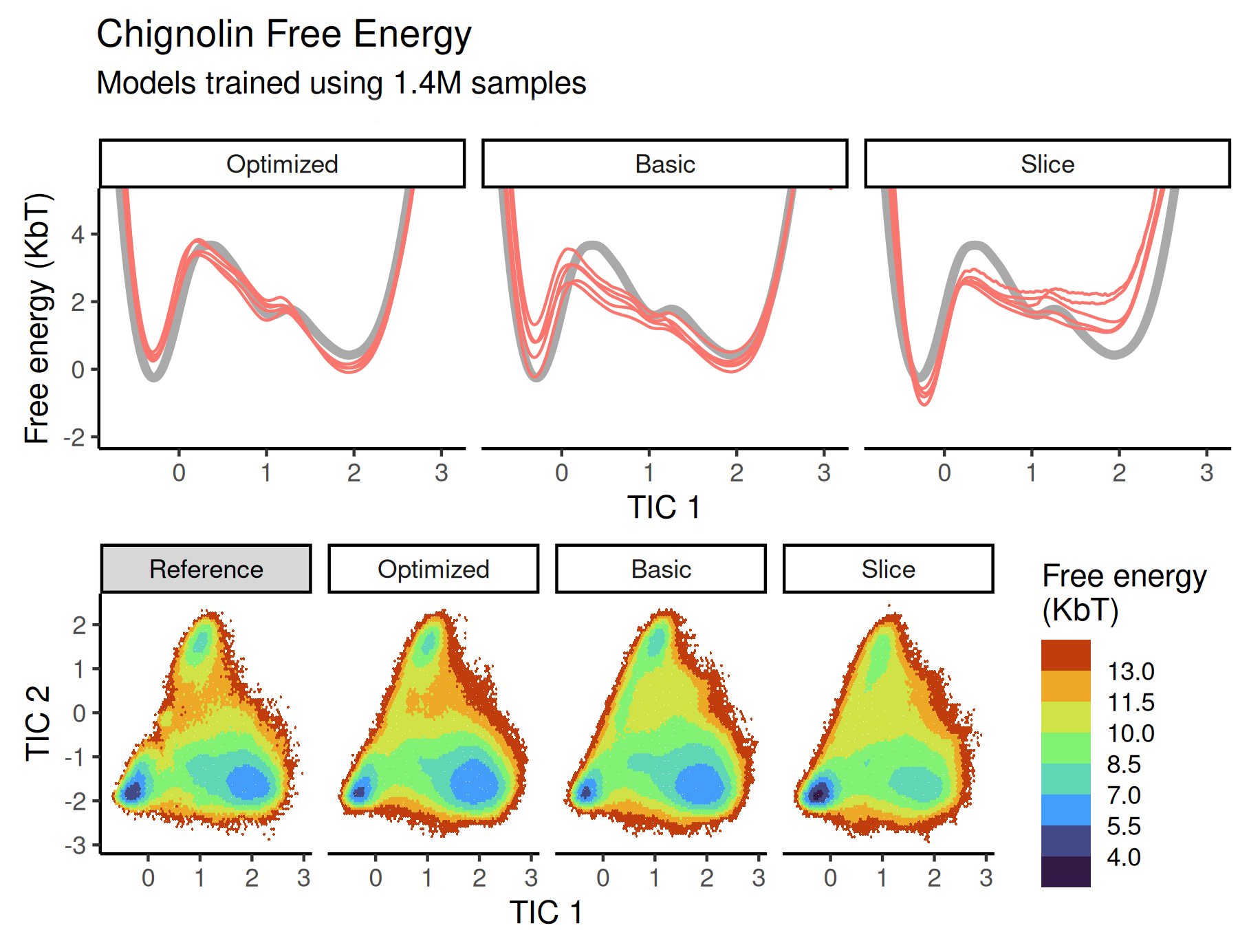}
    \caption{Free energy surfaces calculated for Chignolin. The top row compares surfaces along the slowest TIC: colored lines represent multiple force-fields, each trained using a different subset of the reference trajectory. Grey lines indicate the free energy of the reference trajectory. The bottom row contains surfaces calculated for Chignolin across the two slowest TICs using a single shared subset of the data. Each pane contains data generated using a different force mapping or the reference data for comparison.}
    \label{fig:clnfe}
\end{figure}

\begin{figure}[htbp]
    \centering\includegraphics[width=\linewidth]{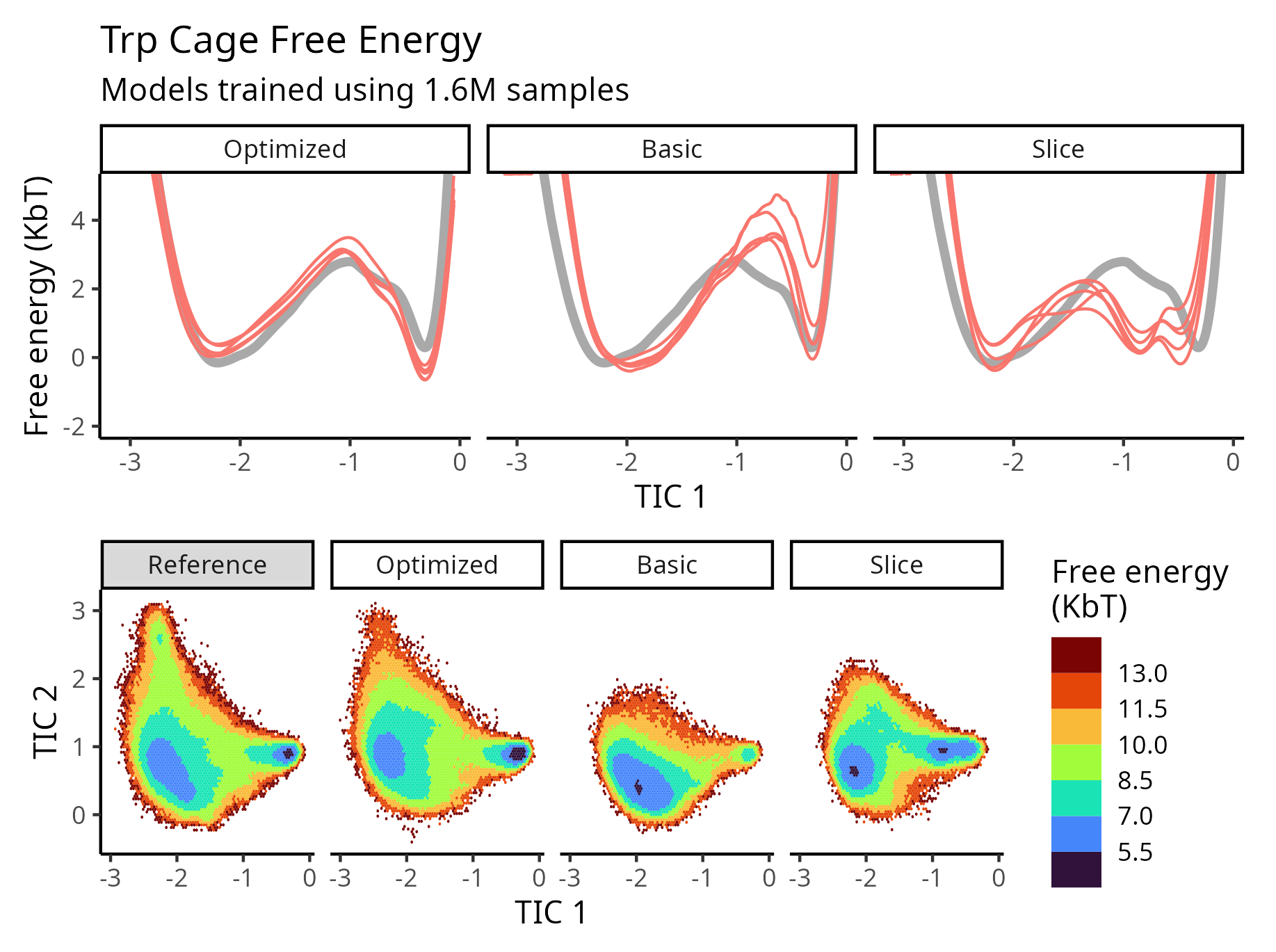}
    \caption{Free energy surfaces calculated for Trp Cage. The top row compares surfaces along the slowest TIC: colored lines represent multiple force-fields, each trained using a different subset of the reference trajectory. Grey lines indicate the free energy of the reference trajectory. The bottom row contains surfaces calculated for Trp Cage across the two slowest TICs using a single shared subset of the data. Each pane contains data generated using a different force mapping or the reference data for comparison.}
    \label{fig:trpfe}
\end{figure}

These measures of errors were calculated for models trained using various subsets of the atomistic data; these subsets were produced using two strategies. First, the effect of reduced dataset size was investigated by striding the atomistic data at a variety of values (see SI). Second, for each stride, the atomistic data was equally partitioned into 5 sections, and 5 models were trained using different subsets of these sections in a strategy similar to cross validation: each model was trained using a different 4/5 of the strided atomistic data. These approaches allow us to study the effect of training set size while quantifying sensitivity to the particular data used. 

\begin{figure}[htbp]
    \centering
    \includegraphics[width=\linewidth]{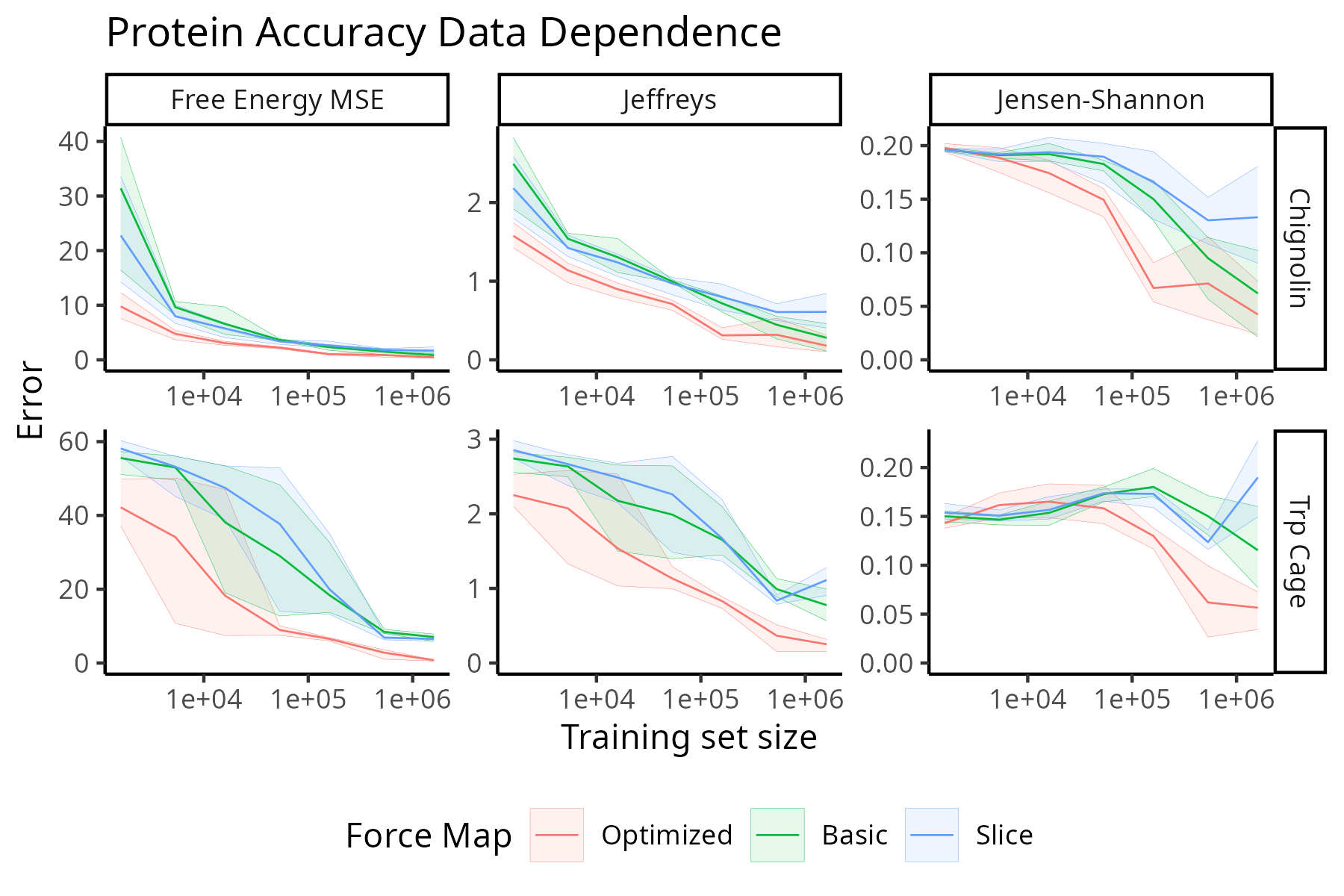}
    \caption{TIC1-TIC2 free-energy error versus training size. Each column specifies an error measure: the mean squared error (MSE) of the free energies, the Jeffreys divergence, or the Jensen-Shannon divergence; each row specifies a protein; and each color represents a force mapping. Each force mapping and training size was investigated by training 5 models on subsets of the reference data (see main text); the mean of the error is plotted as a line, while the maximum and minimum of errors correspond to the bounds of the ribbon.}
    \label{fig:clndatamse}
\end{figure}

The free energy surfaces of the CG models parameterized using large training sets are visualized in Figs. \ref{fig:clnfe} and \ref{fig:trpfe}, and performance of these training procedures as a function of training set size is visualized in Fig.~\ref{fig:clndatamse}. Collectively, the sliced force models exhibit the worst accuracy; their erroneous behavior at large sample size for Trp Cage under the Jeffreys metric is due to spurious states between the folded and unfolded basins (Fig.~\ref{fig:trpfe} and SI). Similar artifacts are seen for large-data Chignolin slice models, as the folded basin is slightly shifted (Fig.~\ref{fig:clnfe} and SI). The behavior in Fig.~\ref{fig:clndatamse} suggests that optimized forces increase efficiency by a factor of approximately 3 over basic forces, each of which avoid the errors produced by the sliced forces. 
Note that, as in the case of the water dimer, optimized forces result in significantly lower force residuals (Figs. \ref{fig:clntrainingcurves} and \ref{fig:trptrainingcurves}). Evaluation of models trained using various force strategies on hold sets using a fixed force aggregation strategy (Table \ref{tab:clnforcerescomp}) demonstrates that optimized-force models result in lower force residuals; however, we note that the success in force prediction and accurate free energy surfaces often have a complex relationship\cite{stocker2022robust,fu2022forces,ricci2022developing}.

Similar to the case of the rigid water dimer, these results strongly suggest that training using invalid slice force mappings introduces artifacts. These errors appear to be resolved by using maps that satisfy the requirements outlined above. While large amounts of training data diminish the advantage of using optimized forces over their basic aggregated counterparts, there does not appear to be a downside to utilizing optimized forces in all situations. Collectively, our results suggest that optimized forces result in less overfitting and lower model variance with regard to both the force residual and free energy surface.

It is important to note that while the expressions in this paper apply to configurational maps which average positions (e.g., a center of mass mapping encompassing each amino acid), these aggregated configurational maps may be less likely to exhibit the problems demonstrated for sliced configurational mappings. This is because the force mappings derived from such aggregation mappings using previously established rules\cite{Noid2008} may satisfy $i)$ and $ii)$ in Eq. \ref{eq:cond1_orthogonal} for typical constraints and incorporate a diverse set of atomistic forces. However, whether such mappings are appropriate for the application depends on other aspects of force-field preparation, such as the imposition of functional forms on bonded force-field contributions. Similarly, we note that future comparisons between force matching results using different configurational mappings should be cognizant of the force mapping used, and that such force-mappings should be reported to facilitate reproduction.

\section{Conclusion}

As machine-learned force-fields become increasingly powerful, the present work paves the way for more efficient optimization of these force-fields. We demonstrate that the selection of force mapping may significantly affect the resulting force-field.
The proposed optimized force mapping schemes reduce overfitting and increase accuracy, robustness, and data-efficiency. The possibility to partly decouple force mapping coefficients from the configurational map may also elevate approaches to optimize configurational mappings alongside the CG potential.\cite{wang2019cgautoencoders} Future work may further exploit the presented variational principle by using position-dependent force mappings and joint optimization of the force map and CG potential.

\begin{acknowledgement}
The authors thank Clark Templeton, Félix Musil, Andrea Gulyas, Iryna Zaporozhets, Atharva Kelkar, Klara Bonneau, David Rosenberger, and Brooke Husic for helpful discussions, additional experiments, and contributions to the code framework. We gratefully
acknowledge funding from the European Commission (Grant No.~ERC CoG 772230
\textquotedblleft ScaleCell\textquotedblright), the International
Max Planck Research School for Biology and Computation (IMPRS\textendash BAC),
the BMBF (Berlin Institute for Learning and Data, BIFOLD), the Berlin
Mathematics center MATH+~(AA1-6, EF1-2) and the Deutsche Forschungsgemeinschaft
DFG (GRK DAEDALUS, SFB1114/A04 and B08). 
C.C. acknowledges funding from the Deutsche Forschungsgemeinschaft DFG (SFB/TRR 186, Project A12; SFB 1114, Projects B03 and A04; SFB 1078, Project C7; and RTG 2433, Project Q05), the National Science Foundation (CHE-1900374, and PHY-2019745), and the Einstein Foundation Berlin (Project 0420815101). \end{acknowledgement}

\begin{suppinfo}
Experimental details of simulations, coarse-grained models, training procedure.

\end{suppinfo}

\providecommand{\latin}[1]{#1}
\makeatletter
\providecommand{\doi}
  {\begingroup\let\do\@makeother\dospecials
  \catcode`\{=1 \catcode`\}=2 \doi@aux}
\providecommand{\doi@aux}[1]{\endgroup\texttt{#1}}
\makeatother
\providecommand*\mcitethebibliography{\thebibliography}
\csname @ifundefined\endcsname{endmcitethebibliography}
  {\let\endmcitethebibliography\endthebibliography}{}

\newpage
\topskip0pt
\vspace*{\fill}
\hspace*{\fill}
\textbf{
-- Supplementary Information --
}
\hspace*{\fill}
\vspace*{\fill}
\newpage

\clearpage
\renewcommand{\theequation}{S\arabic{equation}}
\renewcommand{\thetable}{S\arabic{table}}
\renewcommand{\thefigure}{S\arabic{figure}}
\setcounter{equation}{0}
\setcounter{table}{0}
\setcounter{figure}{0}

\section{Notation}
\begin{itemize}
    \item $\rfg \in \reals^{3\nfg}$: FG coordinates
    \item $\rcg \in \reals^{3\ncg}$: CG coordinates
    \item $\ffg \in \reals^{3\nfg}$: FG forces
    \item $\fcg \in \reals^{3\ncg}$: mapped forces
    \item $\fcg: \reals^{3\nfg} \left(\times\reals^{\# \mathrm{map-parameters}}\right) \rightarrow \reals^{3\ncg}$ mapped forces as a function of FG coordinates
    \item $\constraints: \reals^{3\nfg} \to \reals^{\nconstraints}$: constraints on the FG system
    \item $\cgmap: \reals^{3\nfg} \to \reals^{3\ncg}$: coordinate mapping
    \item $\cgmapmat \in \reals^{3\nfg \times 3\ncg}$: matrix characterizing a linear coordinate mapping via $\cgmapmat \rfg = \rcg$
    \item $\pcgmapmat \in \reals^{\nfg \times \ncg}$: matrix characterizing particle-wise contributions to $\cgmapmat$
    \item $\forcemap: \reals^{3\nfg} \left( \times \reals^{\# \mathrm{map-parameters}}\right) \to \reals^{3\nfg\times 3\ncg}$: force mapping as a function of FG coordinates
    \item $\forcemapmat \in \reals^{3\nfg \times 3\ncg}$: matrix characterizing a ``linear'' force mapping via $\forcemapmat \ffg(\rfg) = \fcg(\rfg)$; sometimes expressed a function of $\paramsmap$
    \item $\pforcemapmat \in \reals^{\nfg \times \ncg}$: matrix characterizing particle-wise contributions to $\forcemapmat$; sometimes expressed a function of $\paramsmap$
    \item $\constraintmat \in \{0,1\}^{* \times \nfg}$: matrix characterizing bond constraints in the atomistic system; dimensions specified in context
    \item $\potfg:\reals^{3\nfg} \to \reals$: FG potential
    \item $\pmf:\reals^{3\ncg} \to \reals$: CG potential of mean force
    \item $\potcg:\reals^{3\ncg}\times\reals^{\# \mathrm{parameters}} \to \reals$: CG potential as a function of CG coordinates
    \item $\raverage{\cdot}$: short-hand for the atomistic ensemble average $\mathbb{E}_{\rfg\sim \exp(-\beta \potfg),\, \constraints(\rfg)= 0} $
    \item $\rRaverage{\cdot}$: short-hand for the conditional average $\mathbb{E}_{\rfg\sim \exp(-\beta \potfg),\, \constraints(\rfg)= 0,\, \cgmap(\rfg) =\rcg}$
    \item $\Raverage{\cdot}$: short-hand for the conditional average $\mathbb{E}_{\rcg\sim \exp(-\beta \pmf)};$ equivalent to $\raverage{\cdot}$
    \item $\paramsmap \in \reals^{*}$: vector describing parameterization of $\forcemap$; length specified in context
    \item $\paramscg \in \reals^{*}$: vector describing parameterization of $\potcg$; length unused
    \item $\paramsmap_I \in \reals^{*}$: vector describing parameterization of $\forcemap$ corresponding to single CG site $I$; length specified in context
    \item $\trajforcemat \in \reals^{3\nfg \times n_t}$: array containing all atomistic forces in a trajectory
    \item $\trajforcematreshape \in \reals^{\nfg \times 3n_t}$: array containing all reorganized atomistic forces in a trajectory
    \item $\nsteps \in \reals^+$: number of frames in a MD trajectory
\end{itemize}

\section{Theoretical considerations}

\subsection{Decomposition of the force matching residual}
This section specifies and discusses the terms in the decomposition of the force matching residual in Eq. \eqref{eq:decomposition}. Following \citeauthor{Wang2019}\cite{Wang2019}, we add and subtract the PMF:
 \begin{align}
        \mscgresidual (\rfg; \paramscg, \paramsmap) 
        &= 
        \left\|
             - \nabla_{\rcg} \potcg(\rcg; \paramscg)
            - \fcg(\rfg; \paramsmap)
        \right\|_2^2
        \\
        &= 
        \|
            \underbrace{
                -\nabla_{\rcg} \potcg(\rcg; \paramscg)
                 + \nabla_{\rcg} \pmf(\rcg)
             }_{=:\bm{\varepsilon}(\rcg; \paramscg)}
             \underbrace{
                - \nabla_{\rcg} \pmf(\rcg)
                - \fcg(\rfg; \paramsmap)
            }_{=:\bm{\zeta}(\rfg; \paramsmap)}
        \|_2^2
        \\
        &= 
        \underbrace{
            \| \bm{\varepsilon}(\rcg; \paramscg) \|_2^2
        }_{ \mathrm{PMF\ error} (\rcg; \paramscg)}
        + 
        \underbrace{
        \| \bm{\zeta}(\rfg; \paramsmap) \|_2^2
        }_{\mathrm{noise} (\rfg; \paramsmap)}
        + 
        \underbrace{
        2 \bm{\varepsilon}(\rcg; \paramscg)^T \bm{\zeta}(\rfg; \paramsmap)
        }_{\mathrm{mixed\ term} (\rfg; \paramscg, \paramsmap)}
        .
        \label{eq:detailed_decomposition}
\end{align}
By definition of valid force maps, Eq. \eqref{eq:valid_map}, the noise
\begin{align}
    &\rRaverage{\bm{\zeta}(\rfg;\paramsmap)}
    = - \nabla_{\rcg}\pmf(\rcg) - \rRaverage{\fcg(\rfg;\paramsmap)}
    = - \nabla_{\rcg}\pmf(\rcg) + \nabla_{\rcg}\pmf(\rcg)= \bm{0}
    \label{eq:zetameanfree1}
    \\
    \mathrm{and} \quad 
    &\raverage{\bm{\zeta}(\rfg;\paramsmap)}
    = \Raverage{\rRaverage{\bm{\zeta}(\rfg;\paramsmap)}}
    = \bm{0}.
\end{align}
As shown in previous work\cite{Noid2008,Wang2019}, inserting Eq. \eqref{eq:zetameanfree1} eliminates the mixed term in the ensemble average of the force residual  (Eq. \eqref{eq:detailed_decomposition}), so that
\begin{equation}
        \raverage{\mscgresidual (\rfg; \paramscg, \paramsmap)}
        = 
        \raverage{\left\|
             - \nabla_{\rcg} \potcg(\rcg; \paramscg)
            - \fcg(\rfg; \paramsmap)
        \right\|_2^2}
        =
        \raverage{\| \bm{\varepsilon}(\rcg; \paramscg) \|_2^2
        }
        + 
        \raverage{\| \bm{\zeta}(\rfg; \paramsmap) \|_2^2}
        \label{eq:mscg_residual}
\end{equation}
It is important to note that $\bm{\zeta}$ depends on $\paramsmap$ and not $\paramscg$, with the opposite holding true for $\bm{\varepsilon}$. 
This has simple but important implications. 
First, minimization of Eq. \eqref{eq:mscg_residual} with respect to $\paramscg$ results in the same minimizer, independent of the force map. 
Second, minimization of Eq. \eqref{eq:mscg_residual} with respect to $\paramsmap$ results in the same minimizer, independent of $\paramscg$. 
 Suppose that $\paramscg_{\mathrm{PMF}}$ and $\paramscg_{0}$ exist such that $\potcg(\rcg; \paramscg_{\mathrm{PMF}}) = \pmf(\rcg)$ and $\potcg(\rcg; \paramscg_0) = 0$. 
 It is then straightforward to see that 
 $\raverage{\mscgresidual (\rfg; \paramscg_{\mathrm{PMF}}, \paramsmap)} 
 = 
 \raverage{\| \bm{\zeta}(\rfg; \paramsmap) \|_2^2}$ 
 and 
 $\raverage{\mscgresidual (\rfg; \paramscg_{\mathrm{0}}, \paramsmap)} 
 = 
 \raverage{\left\|
\fcg(\rfg; \paramsmap)
\right\|_2^2}$. 
As a result, a valid force map which exhibits an optimal 
$\raverage{\left\|
\fcg(\rfg; \paramsmap)
\right\|_2^2}$
has an equivalently optimal $\raverage{\| \bm{\zeta}(\rfg; \paramsmap) \|_2^2}$, i.e.,
\begin{equation}
    \underset{\paramsmap}{\arg\min}  \raverage{\| \bm{\zeta}(\rfg; \paramsmap) \|_2^2} = \underset{\paramsmap}{\arg\min}  \raverage{\| \fcg(\rfg; \paramsmap) \|_2^2}.
    \label{eq:argmin_equivalent}
\end{equation}

\subsection{Variance minimization}
The parameter gradient and its mean are
\begin{align}
  \nabla_{\paramscg} \mscgresidual (\bm{r}; \paramscg, \paramsmap)
    &=  \nabla_{\paramscg}   \| \bm{\varepsilon}(\rcg; \paramscg) \|_2^2
    + 2 \nabla_{\paramscg} \bm{\varepsilon}(\rcg; \paramscg)^T \bm{\zeta}(\rfg; \paramsmap)
    \label{eq:param_grad}
    \\
    \raverage{
        \nabla_{\paramscg} \mscgresidual (\bm{r}; \paramscg, \paramsmap)
    }
    &= \raverage{  \nabla_{\paramscg}   \| \bm{\varepsilon}(\rcg; \paramscg) \|_2^2 },
    \label{eq:param_grad_mean}
\end{align}
meaning that the gradient of the force matching residual is an unbiased estimator of the gradient of the PMF error. 
To facilitate efficient optimization with stochastic gradient-based optimizers, we aim to minimize the
\begin{align*}
    \mathrm{Variance} &=
    \raverage{
        \left\|
        \nabla_{\paramscg} \mscgresidual (\bm{r}; \paramscg, \paramsmap) 
        -
        \raverage{
            \nabla_{\paramscg} \mscgresidual (\bm{r}; \paramscg, \paramsmap)
        }
        \right\|_2^2
    } 
    & \quad \mathrm{(mean\ squared\ deviation)}
    \\
    & = 
    \raverage{
        \left\|
        2 \nabla_{\paramscg} \bm{\varepsilon}(\rcg; \paramscg)^T \bm{\zeta}(\rfg; \paramsmap)
        \right\|_2^2
    }
    & \quad \mathrm{(using\ Eq.\ \eqref{eq:param_grad_mean})}
    \\ 
    & \leq 
    4  \raverage{
      \left\|
       \nabla_{\paramscg} \bm{\varepsilon}(\rcg; \paramscg)
        \right\|_2^2
        \left\|
            \bm{\zeta}(\rfg; \paramsmap)
        \right\|_2^2
        }
    & \quad \mathrm{(consistency\ of\ the\ spectral\ norm)}
    \\
    & \leq 
    4  \left( \sup 
        \left\|
        \nabla_{\paramscg} \bm{\varepsilon}
        \right\|_2^2
        \right)
    \raverage{
        \left\|
            \bm{\zeta}(\rfg; \paramsmap)
        \right\|_2^2
    }
    &\quad \mathrm{(upper\ bound,\ H\ddot{o}lder\ inequality)}
\end{align*}
When we assume Lipschitz continuity of the PMF error with respect to the network parameters, the supremum is finite and we have established the desired relation between the gradient variance and the average noise.

Consequently, we can reduce the gradient variance by minimizing the noise with respect to $\paramsmap.$ As shown in Eq. \eqref{eq:argmin_equivalent}, this is equivalent to minimizing the average magnitude of the mapped forces. Therefore we define the optimal force map through
\begin{equation}
  \paramsmap_{\mathrm{opt}} = \underset{\paramsmap}{\arg\min} \raverage{ \left\| \fcg(\rfg; \paramsmap) \right\|_2^2 }.
\end{equation}

\subsection{Optimization of linear force maps}
When only considering configurational and force maps which do not change as a function of configuration (i.e., linear maps), optimization of $\raverage{\left\|\fcg(\rfg; \paramsmap) \right\|_2^2}$ can be approximated from a reference atomistic trajectory in a straightforward manner using linearly-constrained quadratic programming. 
Restricting force contributions to be particle-specific results in an independent smoothing optimization statement for each CG site $I$ (Eq. \eqref{eq:qp}).
\begin{equation}
\min_{\paramsmap_I} \|
\paramsmap_I \constraintmat \trajforcematreshape
\|^2_2
\label{eq:qp}
\end{equation}
$\trajforcematreshape \in \reals^{\nfg \times 3n_t}$ contains reshaped forces present in a molecular trajectory, $\constraintmat \in \{0,1\}^{|\paramsmap_i|\times\nfg}$ is a sparse matrix representing the molecular constraints present in the atomistic system, and $\paramsmap_I$ is a real vector specifying the force parameters specific to CG site $I$. We note that while the minimization statements may be posed in an unconstrained manner, the provided code performs constrained optimization. Programming constraints related to physical bond constraints are implicitly taken into account via $\constraintmat$ and orthogonality conditions (which determine the CG site under optimization) are specified via explicit linear constraints. A complete formulation of the quadratic programming problem described in Eq. \eqref{eq:qp} is given at the end of this SI. Eq. \eqref{eq:qp} may be reformulated to correspond to a control variate minimization in a straightforward manner through application of quadratic form identities and a null space formulation.

\section{Experimental Details}
\label{sec:experimentaldetails}

\subsection{Water dimer reference systems}
\label{sec:dimersystems}
The potential energy function of the water dimers is defined as two interacting TIP3P molecules \cite{jorgensen1983comparison}. To prevent the waters from drifting apart, all atoms are restrained by an isotropic external harmonic potential around the origin $\potfg_{\mathrm{restraint}}=\frac{k}{2} \|\bm{r}\|^2_2$
with a force constant of $k=$ 3 kJ/mol/nm$^2.$ We investigated two variants of this system, one with flexible and one with rigid internal geometry of the water molecules. An OpenMM implementation is available from the \texttt{WaterCluster} test system in openmmtools \cite{rizzi2019openmmtools}. 

FG reference data was generated by simulating both variants in OpenMM 7.7 \cite{eastman2017openmm} using a Langevin integrator at 300 K and 1 ps$^{-1}$ collision frequency. The time step was 1 fs for the constrained and 0.1 fs for the unconstrained system to ensure that the fast-oscillating covalent hydrogen bonds were sufficiently well resolved. Following 10 ps of equilibration, atomistic coordinates and forces were saved once per picosecond for a simulation time of 50 ns. These simulations resulted in two datasets with $5e4$ data points each. The data are available from the \texttt{github.com/noegroup/bgmol} repository as \texttt{bgmol.datasets.WaterDimerFlexibleTIP3P} and \texttt{WaterDimerRigidTIP3P}.

\subsection{Coarse-grained water dimer model}
\label{sec:dimermodel}
The coarse-graining map for the water dimer was defined as a slicing of oxygens (Fig.~\ref{fig:water}a).
The functional form for the water dimer CG potential was defined as a simple mixture of $n_\mathrm{rbf}=200$ one-dimensional Gaussian radial basis functions (RBF) over the bead-to-bead distance $r_{\mathrm{OO}}$. The RBF centers were fixed as equidistant points in the interval $r_{\mathrm{OO}} \in [0.0, 1.0]$ and the Gaussian standard deviation was defined as $\sigma = 1 / n_\mathrm{rbf}.$ This leaves 200 trainable mixture weights $\paramscg$ to define the CG energy function. The CG energy was implemented in PyTorch\cite{Paszke2019Pytorch} and forces were computed by automatic differentiation.

Note that the this design of the CG potential has an intentional flaw: The actual PMF of the atomistic system over a slice mapping does not just depend on $r_{\mathrm{OO}},$ but also the bead distances to the origin, due to the presence of an external field that removes translation invariance. Therefore, even a perfectly trained CG potential cannot accurately match the PMF over the CG space $\reals^{3\ncg}$. This setting mimics a common situation in higher-dimensional, practical coarse-graining tasks, where the CG model is often not expressive enough to represent the PMF.

\subsection{Water dimer training}
\label{sec:dimertraining}
The water datasets consisting of 50,000 data points were each subsampled using 10 different random seeds to generate 10 independent training runs. For each run 5000 data points were randomly selected from the dataset and set aside for validation. Another 4000 data points were randomly selected and used as training sets. We used such relatively small training sets to emulate the scarcity of data in practical applications.
Training was conducted over 1000 epochs using the Adam optimizer with learning rate 0.01 and a batch size of 128. The model with the best validation loss was selected as the final model for each run. 

\subsection{Chignolin reference systems}
\label{sec:proteinsystems}

The all-atom data for the fast folding variant of Chignolin (CLN025 - YYDPETGTWY) \cite{honda2008crystal}  was the same dataset reported in previous works\cite{Wang2019, Husic2020, wang2021multibody, Chen2021implicit}. We here summarize the simulation details for convenience: Using GPUGRID\cite{buch2010high}
and ACEMD\cite{harvey2009acemd} a cubic simulation box with $40\angstrom$ side lengths was defined and CLN025 was solvated and equilibrated using
TIP3P\cite{jorgensen1983comparison} waters and the CHARMM22*\cite{piana2011robust} force-field. For production runs, a Langevin integrator was used with an integration timestep of $4$ fs
and a friction damping constant of $0.1$ ps$^{-1}$. All hydrogen-heavy atom bonds were holonomically constrained
with $4\times$ heavy hydrogen masses. A MSM-based sampling
approach was used\cite{doerr2014fly}, in which
ten initial simulations were run to generate starting
structures for the remaining shorter adaptive sampling runs. The production simulations
consisted of $3744$ approximately $50$ ns trajectories. This procedure resulted in a total
aggregate time of $187.2$ $\mu$s and $1.8e6$ frames of all-atom coordinates and forces.
Atomistic TICs were created by featurizing the atomistic trajectory using pairwise $\calpha$ distances and a lag time of $4$ ns. Data was not MSM reweighted for training force-fields, but was reweighted when creating reference free energy surfaces; non reweighted data was close to the Boltzmann distribution, with slightly more density in transition areas.

\subsection{Coarse-grained Chignolin model}
\label{sec:proteinmodel}

\emph{210 models were used to create the results in this manuscript. In the visualizations presented, we often show data trained from a single fold for brevity; in these cases, the presented fold was randomly selected.} Following previous works\cite{Wang2019, Husic2020, wang2021multibody}, the CG model of CLN025 was defined by retaining only the $10$ backbone $\calpha$ atoms through a configurational slice mapping. The corresponding force mapping was either the same as the configurational slice mapping, a basic force mapping that incorporated all-atom constraints, or a noise-optimized mapping. The CG force-field was defined as a sum of a prior model and a modified SchNet GNN. The prior model was a Hamiltonian that restrained all sequential $\calpha$-$\calpha$ pseudobonds and $\calpha$-$\calpha$-$\calpha$ pseudoangles using harmonic interactions parameterized through Boltzmann inversion of the all-atom data, as well as  sequential $\calpha$ quadruplet pseudodihedrals via a fifth degree sine/cosine expansion and non-bonded power 6 repulsions. The power 6 nonbonded terms were parametrized by residue-type dependent minimum observed distances in the all-atom dataset. These nonbonded interactions were only applied to CG sites which were not involved in bonds or angle terms together. We note that while our choices of hyperparameters were informed by previous work\cite{Husic2020}, hyperparameters were not scanned over for this publication; furthermore, the hyperparameters used for Trp Cage were borrowed from those used for CLN025 without modification or experimentation.
All modified SchNet models were built and trained using PyTorch Geometric using the hyperparameters found in table \ref{tab:networkhypers}.

\begin{center}
    \begin{table}[ht]
        \centering
        \caption{PyTorch hyperparameters used for network design}
        \label{tab:networkhypers}
        \begin{tabular}[t]{|c|c|} 
            \hline
            Hyperparameter & Value \\ 
            \hline\hline
            Embedding Strategy & Amino acid type (unique termini) \\
            Activation Function & Tanh \\ 
            Distance Cutoff & 0 to 30 $\angstrom$ \\
            Radial Basis Functions & 128 ExpNormal\cite{Unke_Meuwly_2019,tholke2022equivariant} \\
            Num Filters & 128  \\
            Filter Cutoff & Cosine Cutoff\cite{Schutt_Kessel_Gastegger_Nicoli_Tkatchenko_Muller_2019,tholke2022equivariant} \\
            Interaction Blocks & 2  \\
            Terminal Network Layer Widths & [128,64] \\
            \hline
        \end{tabular}
    \end{table}
\end{center}

\subsection{Chignolin model training}
\label{sec:proteintraining}

\begin{figure}[htbp]
    \centering\includegraphics[width=\linewidth]{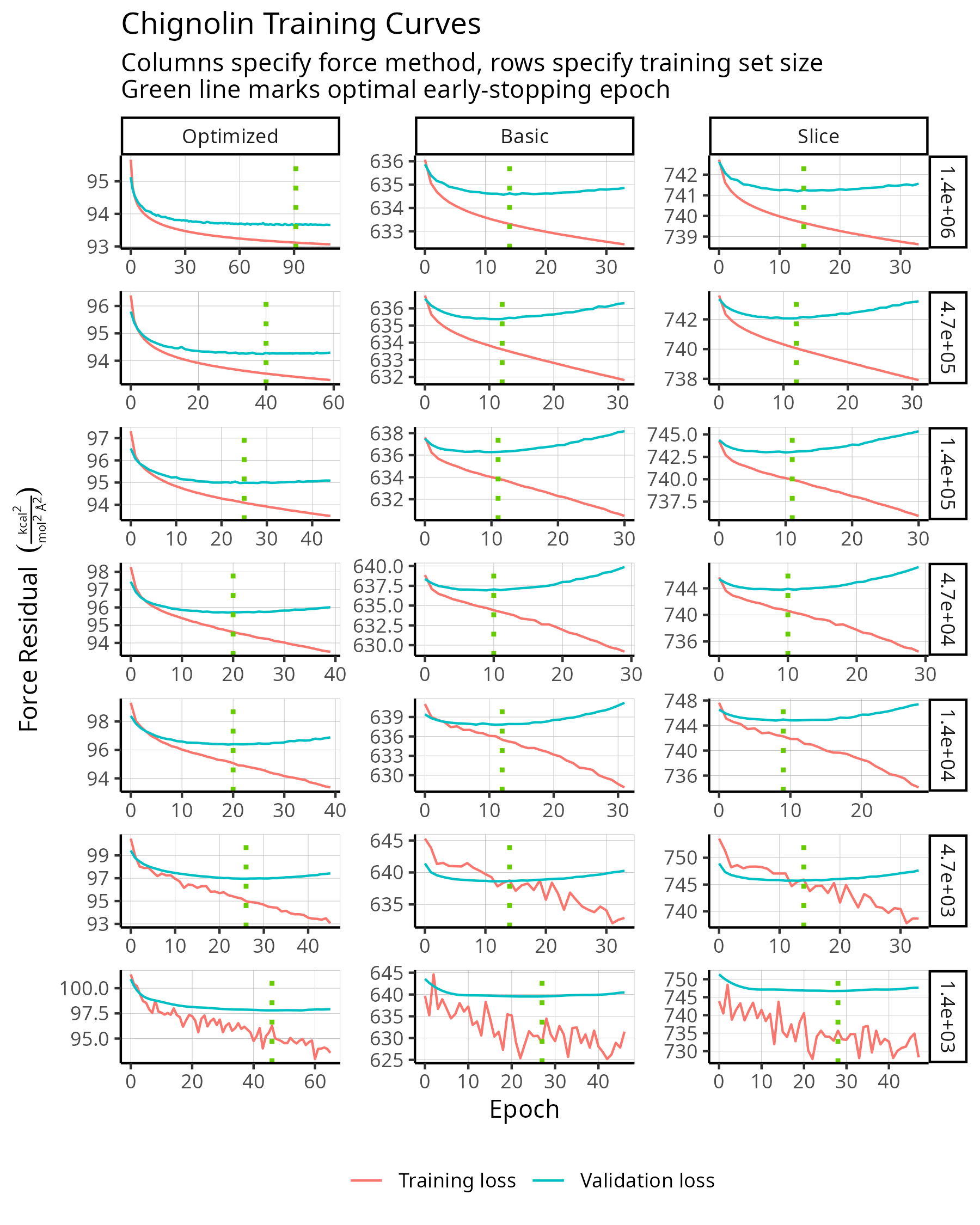}
    \caption{Evolution of force residual for Chignolin during training for models trained on a single fold of data. Note that the axes are not shared between panels.}
    \label{fig:clntrainingcurves}
\end{figure}

Network training was done using PyTorch Lightning. Optimal models were selected from the epoch with the lowest validation error; evolution of the force residual during training is visualized in Fig.~\ref{fig:clntrainingcurves}. Table \ref{tab:traininghypers} summarizes training hyperparameters. Unless otherwise specified, default options were used for optimizers and weight and bias initializations.
\begin{center}
    \begin{table}[ht]
        \centering
        \caption{PyTorch hyperparameters used for network optimization}
        \label{tab:traininghypers}
        \begin{tabular}[t]{|c|c|} 
            \hline
            Hyperparameter & Value \\ 
            \hline\hline
            Optimizer & Adam \\
            Learning Rate & $1e{-4}$ \\
            Batch Size & 512 \\
            GPU & GeForce RTX 1080Ti \\
            \hline
        \end{tabular}
    \end{table}
\end{center}

\subsection{Chignolin model validation}

\begin{figure}[htbp]
    \centering
    \includegraphics[width=\linewidth]{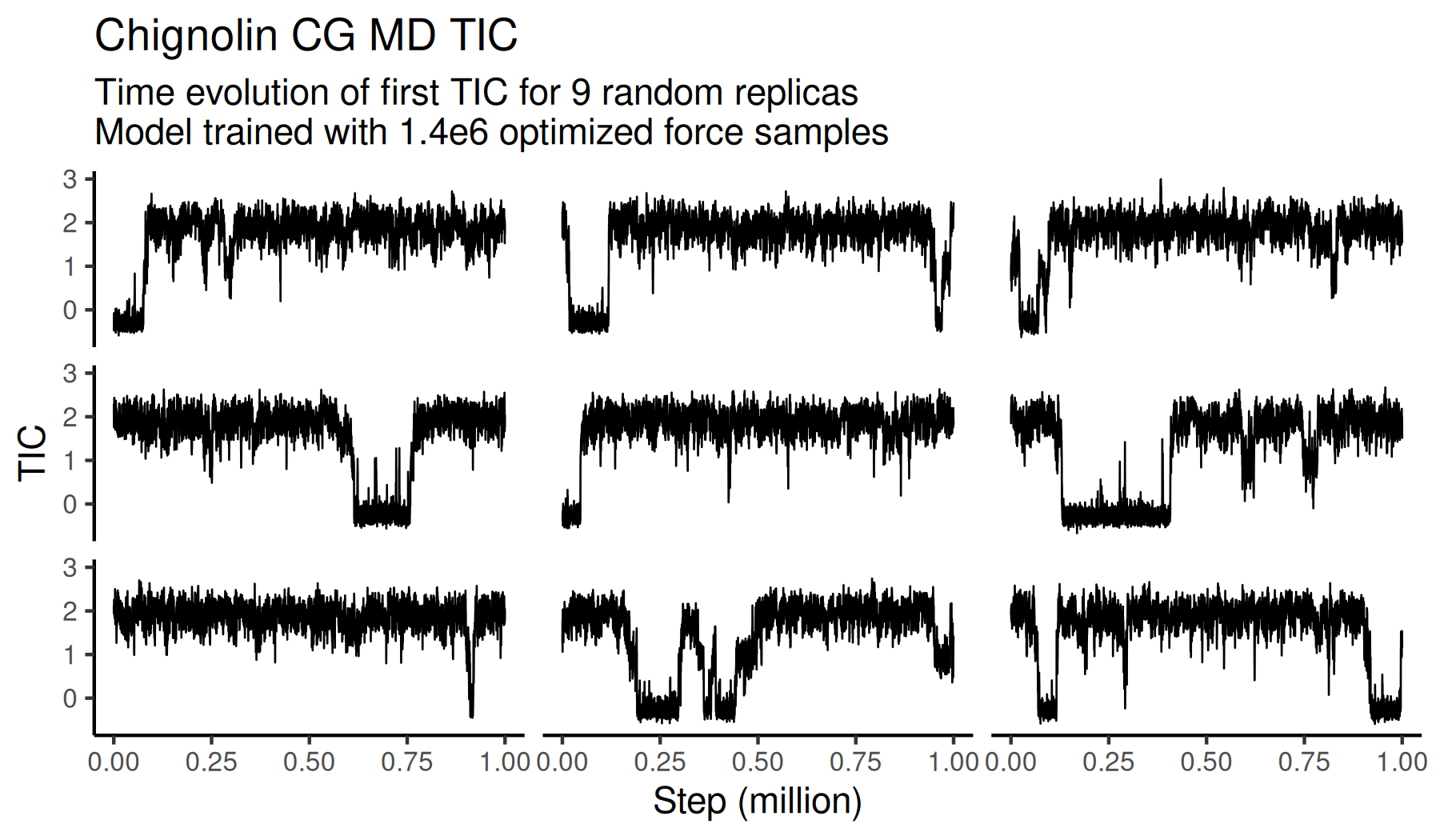}
    \caption{Time evolution of the first TIC evaluated on 9 randomly selected CG MD trajectories. The displayed Chignolin CG model was trained using optimized forces on $1.4e6$ reference frames. Similar (or higher) levels of recrossing were observed for all CG models.}
    \label{fig:clncgmdconvergence}
\end{figure}

 Each Chignolin model was characterized by performing 100 replicas of MD, each for $1e6$ steps, with a Langevin integrator using a friction coefficient of 1 ps$^{-1}$ and a 2 fs timestep; these simulations were seeded from configurations randomly selected from the reference atomistic trajectory. The resulting CG MD was observed to be converged based on time evolution of the leading atomistic TIC (e.g., Fig.~\ref{fig:clncgmdconvergence}); the first $4e5$ frames were discarded before analysis to remove bias related to initial conditions. A small number of these simulations ($<0.1\%$), particularly those trained with smaller datasets, exhibited integration instability. However, upon reinitialization of the MD procedure these problematic simulations successfully completed. Furthermore, note that while previous work\cite{Husic2020} has averaged over the output of multiple models to improve accuracy, averaging was not performed for any model presented in this manuscript. Visulizations were producing using R and associated packages\cite{R,wickham2011ggplot2,dowle2019package}.

\begin{figure}[htbp]
    \centering\includegraphics[scale=1.4]{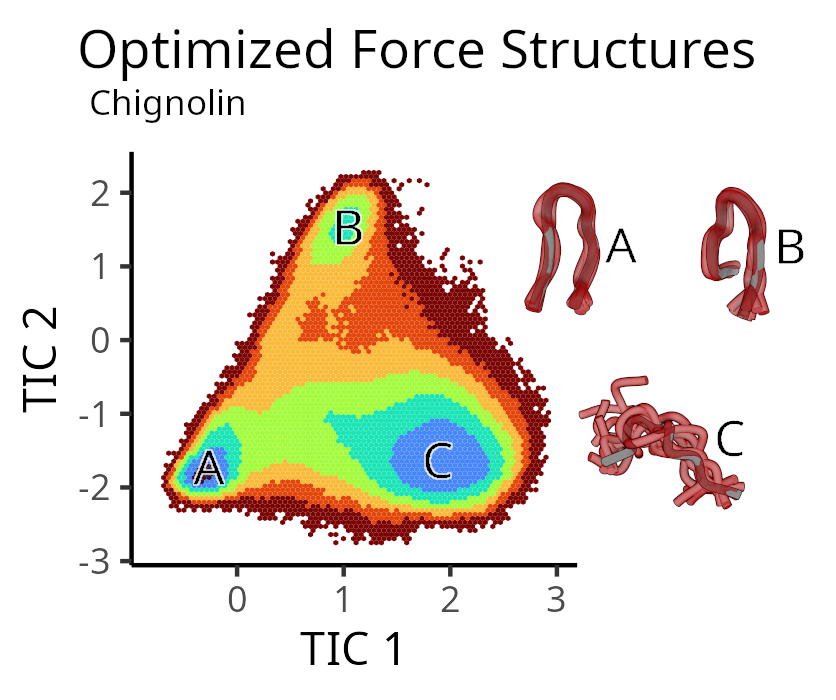}
    \caption{Visualization of structures typical to each basin in the optimized force Chignolin model trained using $1.4e6$ samples. Red structures characterize the CG model ensemble, while grey structures are drawn from the reference trajectory at the same location in TIC space. Note that the reference atomistic basins are also characterized by ensembles, but only one structure is shown for visual clarity.}
    \label{fig:clnoptimfestructures}
\end{figure}

Based on visual inspection of structures and free energy surfaces, the trained models of Chignolin did not typically produce spurious or nonphysical structures. Instead, inaccurate models overaccentuated various basins (e.g., the folded basin). Exceptions to this observation were the slice models of Chignolin at large training set sizes; this deviation is described in the next paragraph. A visualization of the structures produced by the $1.4e6$-sample optimized force model superimposed on representative structures from the reference trajectory is shown in Fig.~\ref{fig:clnoptimfestructures}. For additional descriptions of the structures associated with each basin, see previous work\cite{Wang2019,Husic2020}. Due to the lack of distortion and spurious basins, structures from other models schemes are visually similar to those in Fig.~\ref{fig:clnoptimfestructures} and thus omitted for brevity.

\begin{figure}[htbp]
    \centering\includegraphics[width=\linewidth]{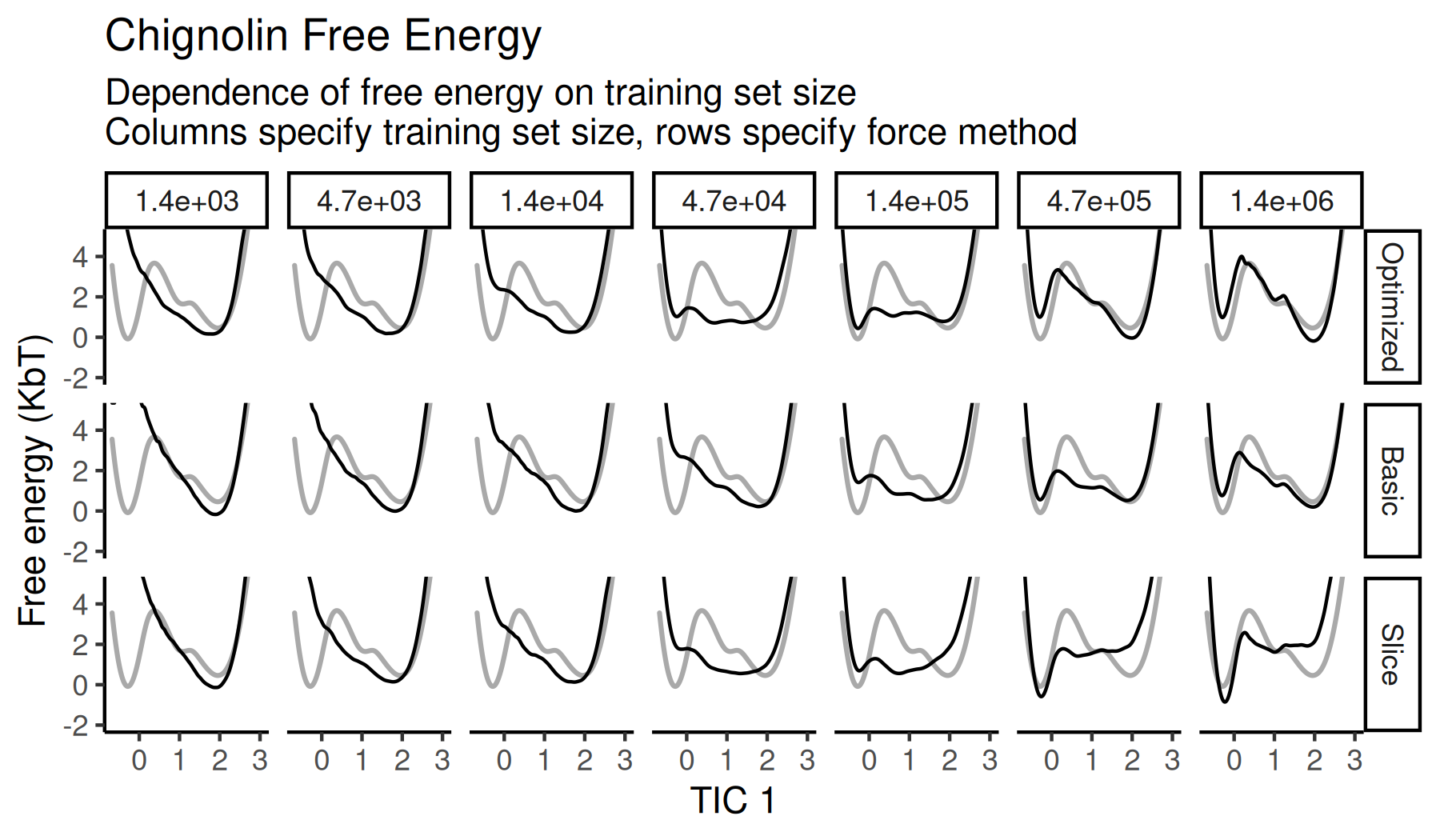}
    \caption{TIC 1 free energy surfaces calculated for Chignolin as a function of training set size for select subsets of the data. Rows specify the force aggregation method, while columns specify the training size. Larger training data sizes produce more accurate models, with the exception of the invalid slice forces. Black lines correspond to models, while grey lines correspond to all-atom reference data.}
    \label{fig:clndatadep}
\end{figure}

\begin{figure}[htbp]
    \centering\includegraphics[width=\linewidth]{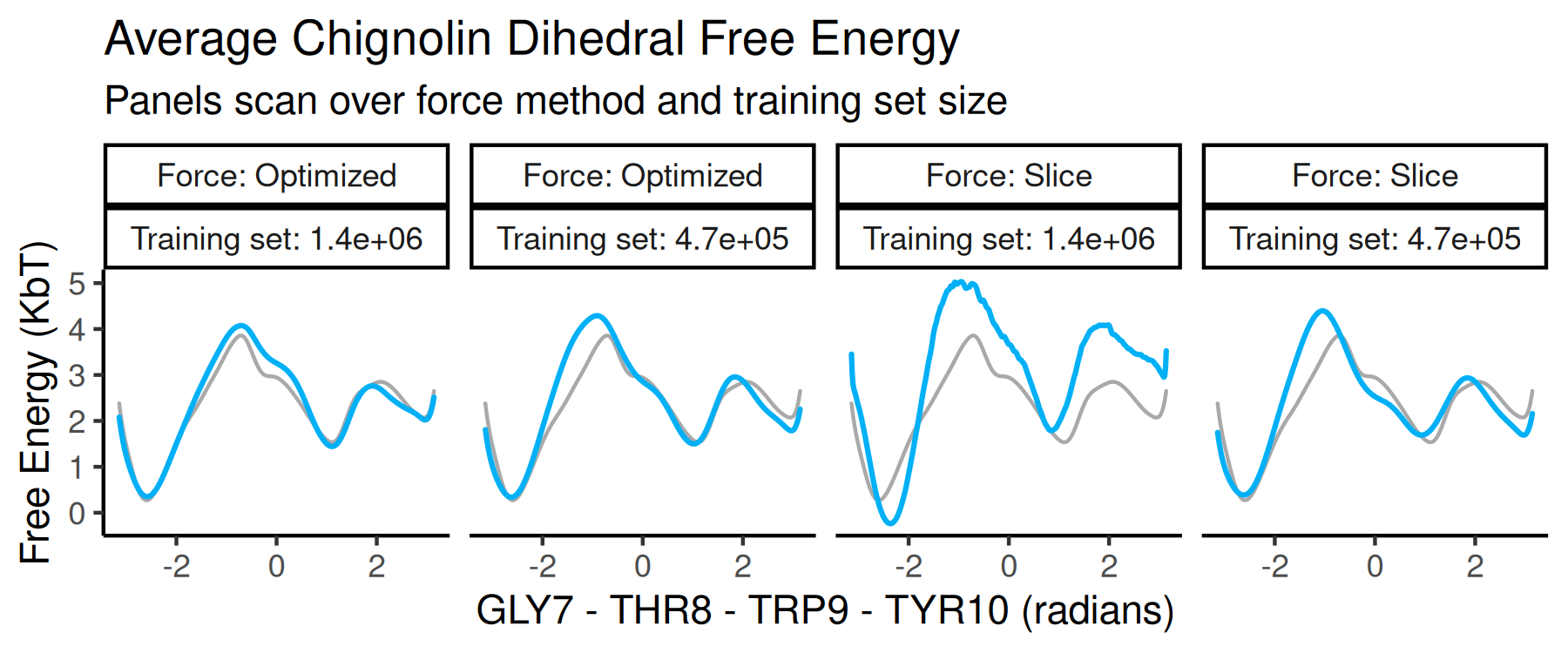}
    \caption{Free energy surfaces calculated along the terminal $\calpha$ pseudodihedral between GLY7, THR8, TRP9, and TYR10. At high training set sizes slice forces produce a distorted dihedral distribution. No such anomaly is present in the optimized forces for any training size (nor basic forces, not shown). Smaller training data sizes do not cause erroneous dihedral distributions.}
    \label{fig:dihedralerror}
\end{figure}

At lower data sizes, free energy surfaces generally appear smoother, with the correct folded and misfolded structures appearing starting at $4.7e4$ training samples for the optimized force model and $1.4e5$ samples for the slice and basic force models. We provide a visualization of this effect for a single model for each force-size combination in Fig.~\ref{fig:clndatadep}. However, as seen in the error trends and free energy surfaces presented in the main text, the CLN025 models trained on slice force mappings display a slight systematic shift in the location of the folded minima along TIC 1 free energy curves with respect to the all-atom reference at large training data sizes. This small shift manifests as a stabilization of slightly shifted dihedral angles in the optimal folded CG structure, with the largest shifts associated with terminal dihedrals and the dihedrals directly preceding and following $\calpha$-GLY7. While this is not apparent from visualization of CG folded structures, it can be  detected by inspecting the above-mentioned dihedral free energy surfaces (Fig.~\ref{fig:dihedralerror}). 

For the reported numerical accuracy trends (e.g, MSE), trajectories were projected onto the first two atomistic TICs, binned, and then compared to the MSM-reweighted reference atomistic data projected and discretized in the same manner. Discretization was performed using 200 equally sized bins spaced from -4 to 4 Angstroms along each TIC axis (resulting in 40000 bins total). Bins which did not have any population were assigned a density of $1e^{-10}$. While exact values of the free energy error metrics changed as a function of bin size, accuracy trends remain stable over a large range of bin resolutions. 

Hold-out force residuals for models trained using optimized forces were consistency slightly lower than those trained using basic or sliced forces (e.g., table \ref{tab:clnforcerescomp}) when using a fixed hold-out force aggregation strategy. Due to the small difference in holdout force residual values, such comparisons require that the the hold out set be held constant and mapped using a single shared force map. While the low magnitude of the observed difference in force residuals may be surprising, we note that relationship between free energy surface quality and force residual is tenuous\cite{fu2022forces}, and that previously reported force residuals\cite{Husic2020} for CLN025 have used the invalid slice force map, impeding understanding whether the observed difference in force residuals is significant. We leave a systematic analysis of these effects to a future study.

\begin{table}[ht]
    \centering
    \renewcommand\arraystretch{1.2}
    \begin{tabular}[t]{|c|c|} 
        \hline
        Training force type & Basic agg. force score $\left(\frac{kcal}{mol \cdot \angstrom}\right)^2$\\
        \hline\hline
        Slice & 633.922 \\
        Basic Agg. & 633.231 \\
        Optimized Agg. & 633.029 \\
        \hline
    \end{tabular}
    \caption{Comparison of hold-out residuals for Chignolin. Training used specified forces while hold out evaluation used basic aggregation.}
    \label{tab:clnforcerescomp}       
\end{table}

\subsection{Trp Cage reference systems}

\begin{figure}[htbp]
    \centering
    \includegraphics[scale=0.15]{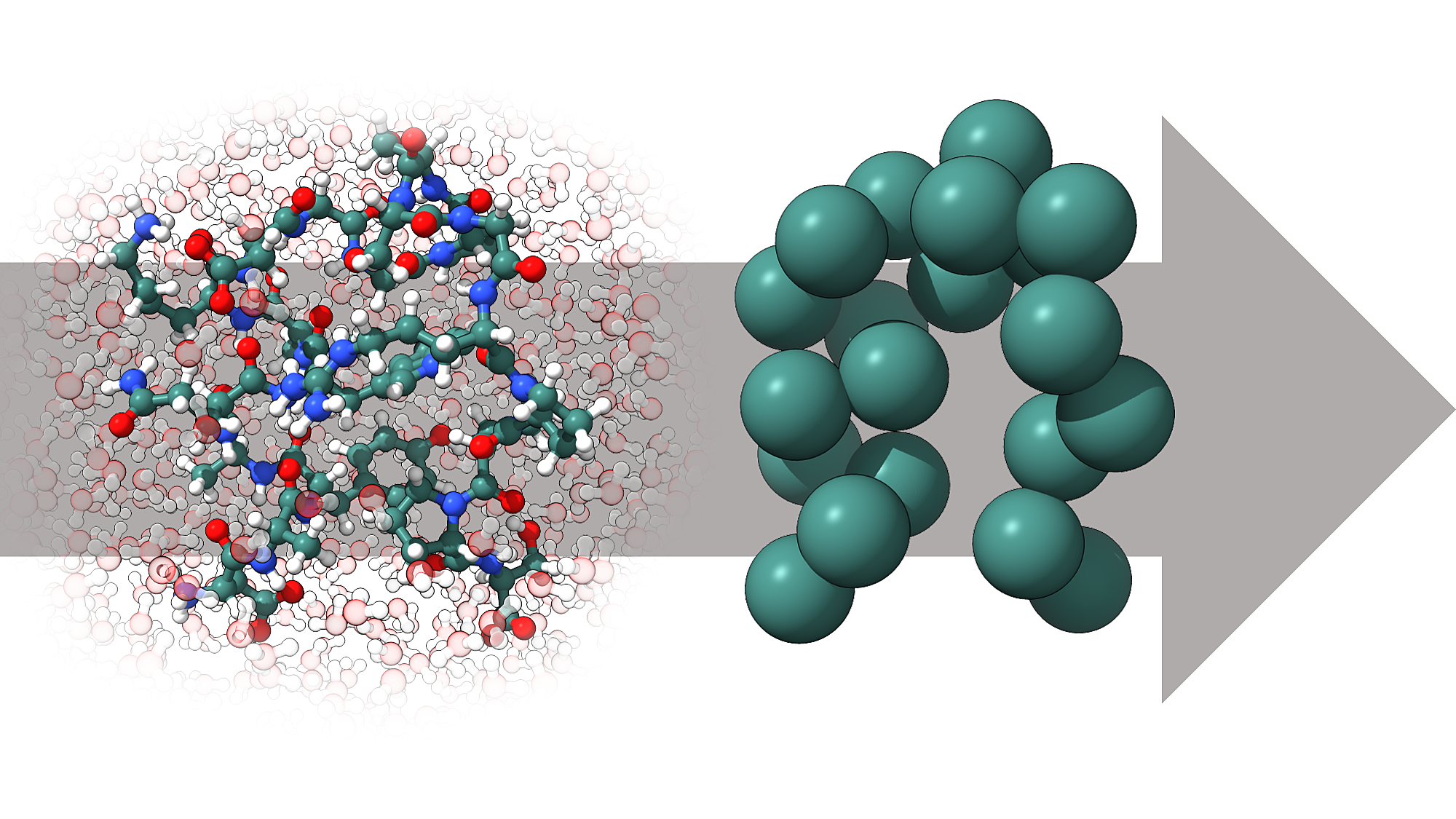}
    \caption{Visualization of the configurational CG mapping used to model Trp Cage. The solvated atomistic resolution used for the reference simulations is shown on the left, while the CG representation (which preserves only $\calpha$s) is shown on the right.}
    \label{fig:trpcg}
\end{figure}

Atomistic data for Trp Cage (DAYAQWLKDGGPSSGRPPPS)\cite{barua2008trp} was generated in a similar manner to that for CLN025; see previous work\cite{majewski2022machine} for a full description. We report the details which differed than those of CLN025 for convenience. 
The production simulations consisted of $3940$ approximately $50$ ns trajectories. This procedure resulted in a aggregate time of $197.3$ $\mu$s and $2e6$ frames of all-atom coordinates and forces. TICs were created by featurizing the atomistic trajectory using a lag time of $12$ ns and pairwise atomic distances; however, unlike Chignolin, the 4 C-terminal residues were omitted from the distance featurization. This is due to an unexpected slow degree of freedom present in the C-terminus of the protein which occurs on a slower timescale than folding. Omission allows the generated TIC coordinates to capture the intuitively important folding states present in the atomistic trajectory. Note that reference atomistic data was not reweighted using the MSM for training, but was reweighted for creating reference free energy surfaces. A visualization of the CG representation is found in Fig.~\ref{fig:trpcg}.

\subsection{Coarse-grained Trp Cage model}
CG models of Trp Cage were trained using identical hyperparameters as those for Chignolin. 

\subsection{Trp Cage model training}
\begin{figure}[htbp]
    \centering\includegraphics[width=\linewidth]{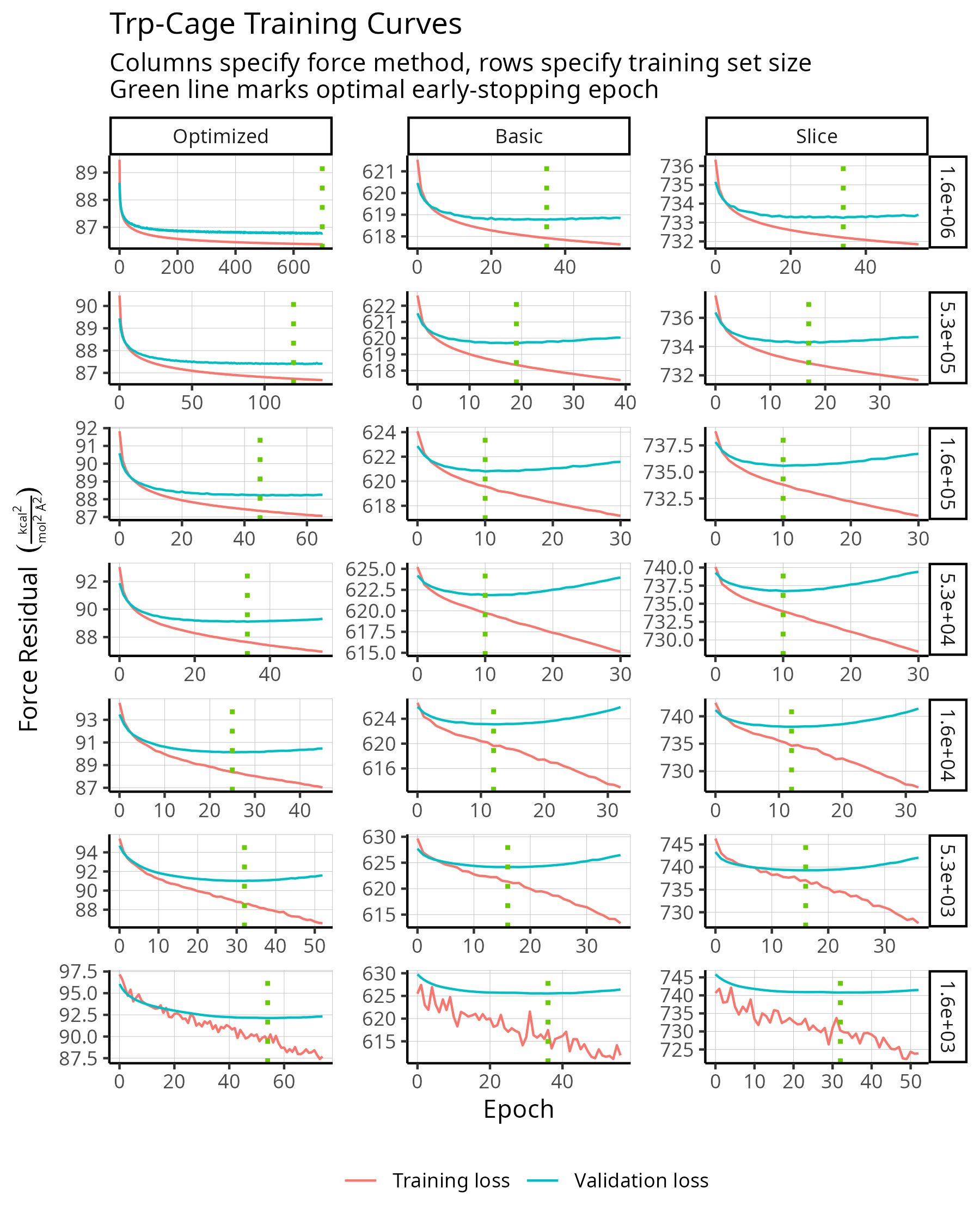}
    \caption{Evolution of force residual for a subset of Trp Cage models during training. Note that axes are not shared between panels.}
    \label{fig:trptrainingcurves}
\end{figure}

The procedure used to train the Trp Cage models was nearly identical to that used for CLN025. However, in the case of Trp Cage, models which were trained on the largest data sizes ($1.6e6$ frames) with optimized forces saw no uptick in the force validation loss
(Fig.~\ref{fig:trptrainingcurves}). As a result, model training was terminated at 700 epochs. 

\subsection{Trp Cage model validation}

\begin{figure}[htbp]
    \centering\includegraphics[scale=1.4]{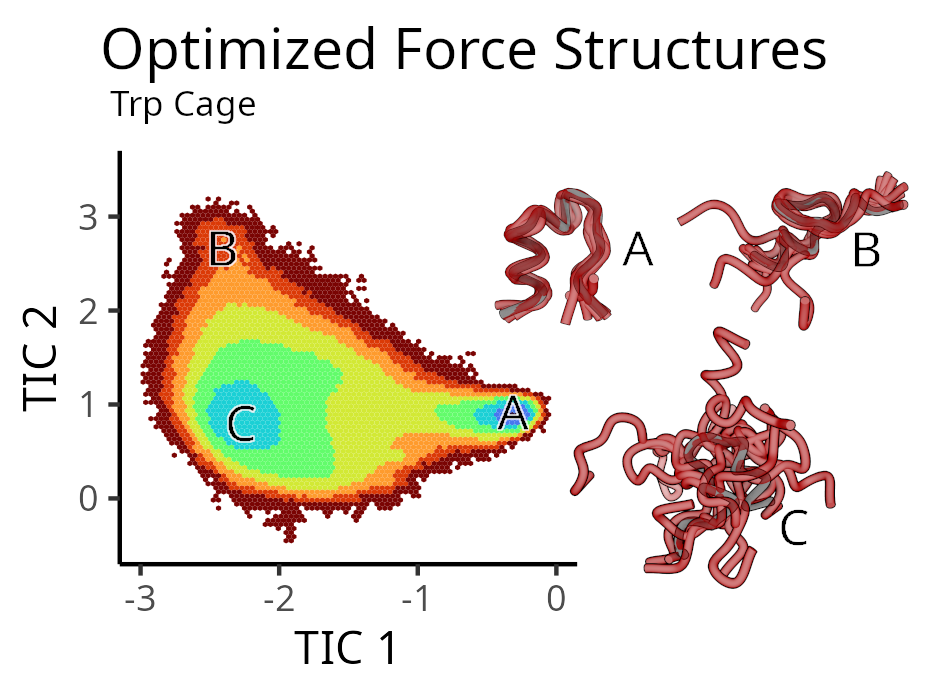}
    \caption{Visualization of structures typical to each basin in the optimized force Trp Cage model trained using $1.6e6$ samples. Red structures characterize the CG model ensemble, while grey structures are drawn from the reference trajectory at the same location in TIC space. Note that the reference atomistic basins were also characterized by ensembles, but only one structure is shown for clarity.}
    \label{fig:trpoptimfestructures}
\end{figure}

\begin{figure}[htbp]
    \centering\includegraphics[scale=1.4]{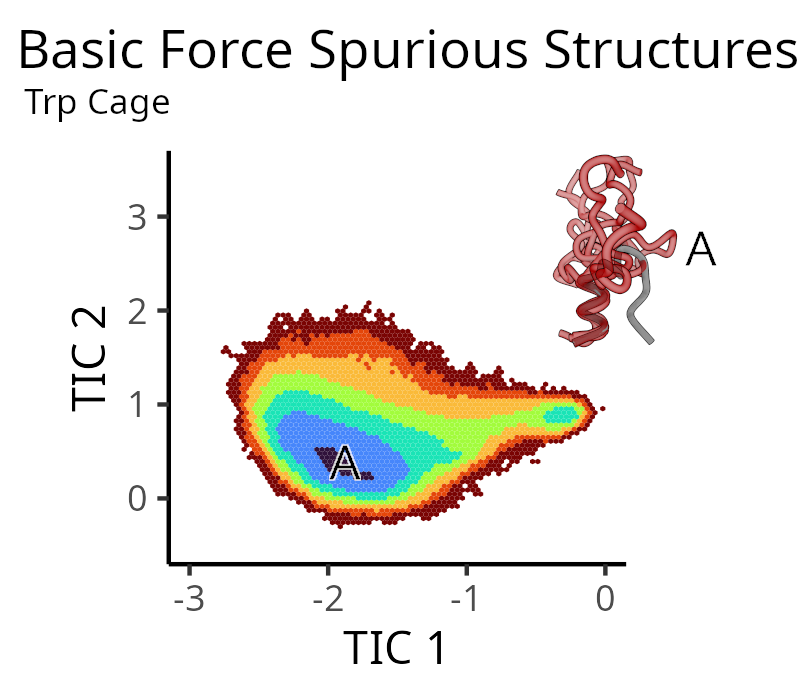}
    \caption{Structures typical to the \emph{incorrect} basin in the basic force Trp Cage model trained using $1.6e6$ samples. Red structures characterize the CG model ensemble, while the grey structure is that of reference folded basin. Model structures exhibit a reliably formed helix, but the rest of the protein is unfolded.}
    \label{fig:trpbasicfestructures}
\end{figure}

\begin{figure}[htbp]
    \centering\includegraphics[scale=1.4]{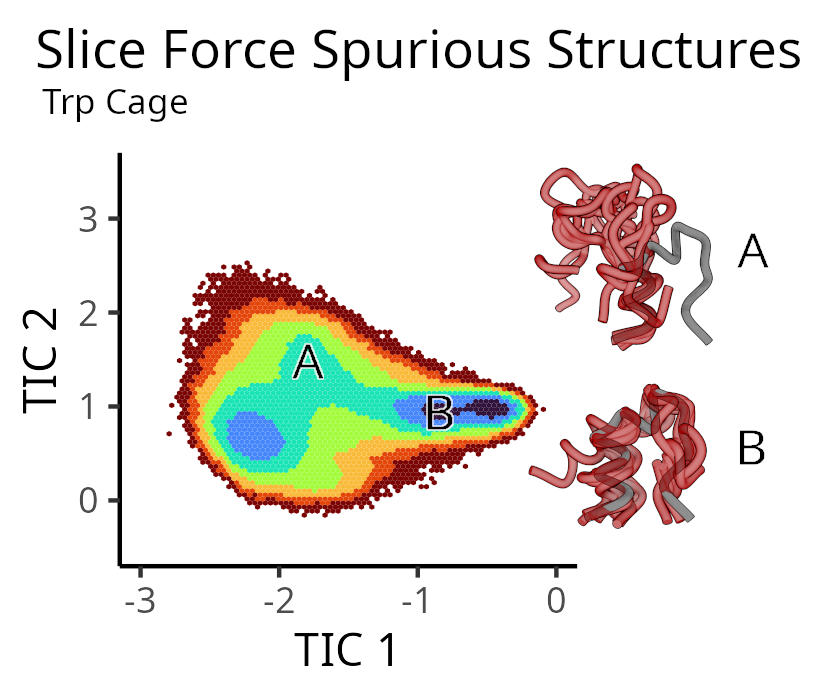}
    \caption{Structures typical to each \emph{incorrect} basin in the slice force Trp Cage model trained using $1.6e6$ samples. Red structures characterize the CG model ensemble, while the grey structures are that of reference folded basin. Model structures in basin A contain a partial helix. The difference between the states of the basic force model and slice basin A are difficult to distinguish, but have distinct disordered ensembles. Basin B exhibits a somewhat folded structure, but with distortions throughout.}
    \label{fig:trpslicefestructures}
\end{figure}

\begin{figure}[htbp]
    \centering
    \includegraphics[width=\linewidth]{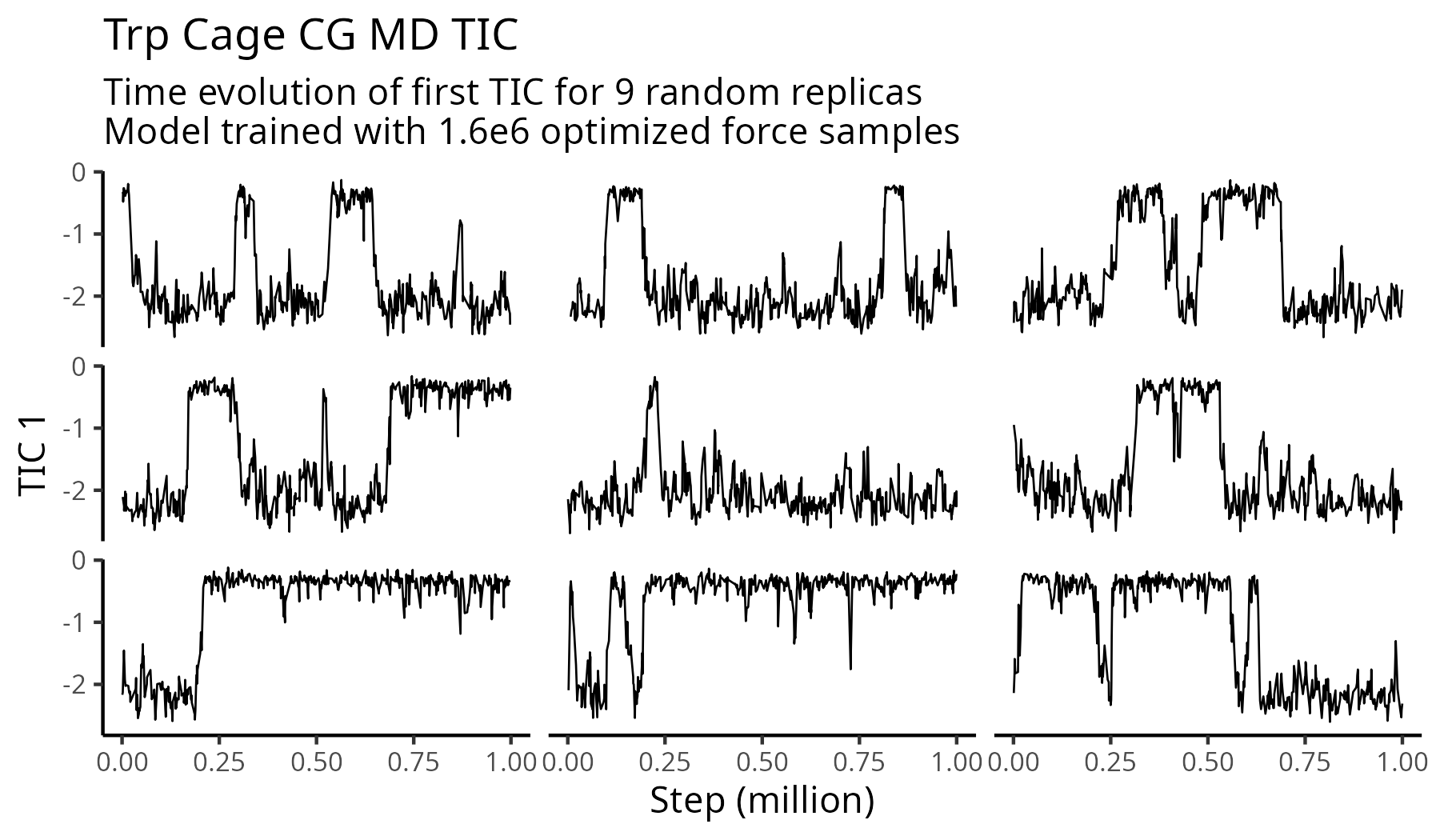}
    \caption{Time evolution of the first TIC evaluated on 9 randomly selected CG MD trajectories. The displayed Trp Cage CG model was trained using optimized forces on $1.4e6$ reference frames. Similar (or higher) levels of recrossing were observed for all CG models.}
    \label{fig:trpcgmdconvergence}
\end{figure}

\begin{figure}[htbp]
    \centering\includegraphics[width=\linewidth]{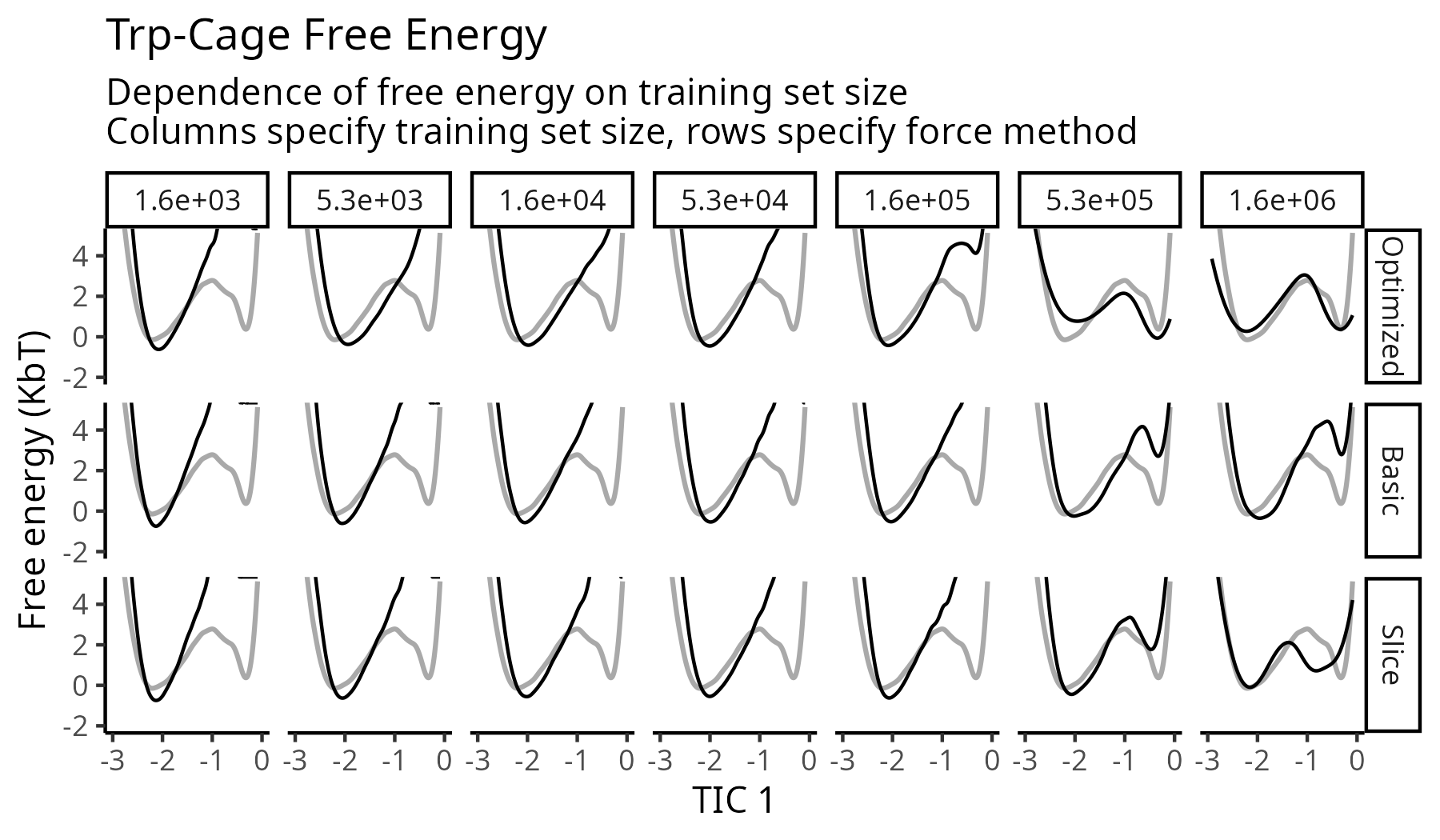}
    \caption{TIC 1 free energy surfaces calculated for Trp Cage as a function of training set size for select subsets of the data. Rows specify the force aggregation method, while columns specify the training size. Larger training data sizes produce more accurate models, with the exception of the invalid slice forces. Black lines correspond to models, while grey lines correspond to all-atom reference data.
    }
    \label{fig:trpdatadep}
\end{figure}

Models were characterized using the same MD setup as for Chignolin; the time evolution of the leading atomistic TIC over a sample of CG MD trajectories is found in Fig.~\ref{fig:trpcgmdconvergence}. A visual comparison of reference structures and the structures found in a $1.6e6$-sample optimized force model is provided in Fig.~\ref{fig:trpoptimfestructures}; while basin depths exhibited deviations, the shape of the basins were reasonably accurate. Basin A corresponds to the folded state, basin B contains a misfolded state, and basin C describes the unfolded state. The misfolded state is primarily characterized by a partially formed helix, an inverted sheet and PRO13 hairpin turn, and an absence of the closing of the tryptophan cage. Similarly accurate folded structures are found in the basic force folded basins; the shifted minima of the basic unfolded basin is due incorrect residual helix structure without correct tertiary packing or turn formation (Fig.~\ref{fig:trpbasicfestructures}). The slice model exhibited two spurious basins (Fig.~\ref{fig:trpslicefestructures}). The one closest to the folded state is characterized by a distorted helix and turn, but is difficult to attribute to any simple error involving one or few residues. The other is characterized by a partially formed helix but no correct turn. We note that characterization of spurious model states is challenging, as interpretation of arbitrary structures, especially at the $\calpha$ resolution, is difficult; however, it is critical to realize that basins which are shifted in TIC space reliably exhibit some systematic deviation from the reference data. A visualization of TIC 1 free energy surfaces as a function of training set size and force strategy is presented in Fig.~\ref{fig:trpdatadep}. Divergences and MSE errors for Trp Cage were calculated using nearly the same methodology as was used for CLN025, except that the TICs were discretized using $(4, 1)$ and $(-1, 4)$ as bounds.


\section{Smoothing forces using quadratic programming}

As mentioned in the appendix, under certain conditions it is possible to numerically minimize an upper bound on the variance of the gradient estimator by optimizing $\raverage{\left\|\fcg(\rfg; \paramsmap)\right\|_2^2}$. We here explicitly construct the numerically viable optimization statement in Eq. \eqref{eq:qp} and discuss its implications. The results in this section apply to linear configurational and force maps, with additional limitations being that we only consider constrained bonds and that the force map is defined particle-wise; this last term we will define below.
We begin by stating the optimization residual for a linear force map (Eq. \eqref{eq:linear_force_residual})
\begin{equation}
\raverage{\left\|\fcg(\rfg; \paramsmap)\right\|_2^2}
=
\raverage{\left\|\forcemapmat(\paramsmap)\aaforce(\rfg)\right\|_2^2}
\label{eq:linear_force_residual}
\end{equation}
where we have introduced $\forcemapmat$, a real-valued matrix characterizing the ``linear'' (i.e., configuration independent) force map with shape $3N\times3n$. 

Expansion of the norm and moving the sum outside the ensemble average results in Eq. \eqref{eq:qp_outer_sum}, where we have used $I$ and $d$ to iterate over rows of $\forcemapmat$ by specifying CG index and dimension, respectively.
\begin{equation}
\sum_{I,d}
\raverage{
\left[
\forcemapmat({\paramsmap})_{I,d} \cdot \aaforce(\rfg)
\right]^2
}
\label{eq:qp_outer_sum}
\end{equation}
If molecular constraints are not present and the parameterization is chosen appropriately, the summands in Eq. \eqref{eq:qp_outer_sum} may be optimized independently under appropriate constraints: each particle-wise force map maintain orthogonality relations to the \emph{configurational} (not force) mapping (Eq. \eqref{eq:cond2_basis}). However, the inclusion of molecular constraints, as well as the particle-wise decomposition described in the next paragraph, necessitate further modification to Eq. \eqref{eq:qp_outer_sum}.

$\forcemapmat$ specifies how each \emph{Cartesian coordinate} of each atom contributes to each Cartesian coordinate of each CG site. We reduce this flexibility (and dimensionality of the resulting optimization) by only specifying weights for each particle pair; e.g., atom $2$'s $x$ component contributes to the fluctuating force of CG site $5$'s $x$ component and equal amount as do their $y$ components. This can be expressed by a particle-wise force matrix of size $N\times n$, which we denote $\pforcemapmat$, and is referred to as using particle-wise force contributions. Symbolically, this is expressed as Eq. \eqref{eq:pforcemapmat}, where we have used $\outerprod$ to denote the outer matrix product.
\begin{equation}
\pforcemapmat({\paramsmap})\outerprod \ident_3 := \forcemapmat({\paramsmap})
\label{eq:pforcemapmat}
\end{equation}
As our parameterized force contribution coefficients are now shared along all Cartesian coordinates specific to our atoms and CG particles, Eq. \eqref{eq:qp_outer_sum} may now only be separated along $I$, resulting in $N$ independent suboptimizations (Eq. \eqref{eq:qp_sub_opt}), where we have reordered indices to reflect a parameterization allows per particle optimization.
\begin{equation}
\left[
\sum_{d}
\raverage{
\left[
\forcemapmat_{I,d}({\paramsmap_I}) \cdot \aaforce(\rfg)
\right]^2
}
\right]_I
\label{eq:qp_sub_opt}
\end{equation}

The next design constraint that must be satisfied is compatibility with respect to atomistic constraints. Here, we consider only quadratic bond constraints; in this case, if two atoms are connected by a bond constraint, it suffices to ensure that the force contributions these atoms make to any CG site are equal. This is enforced through our parameterization of $\pforcemapmat$ (Eq. \eqref{eq:qp_force_map_param}), 
\begin{equation}
\pforcemapmat_{I} ({\paramsmap_I}) := \paramsmap_I \constraintmat 
\label{eq:qp_force_map_param}
\end{equation}
where we have stated our definition row-wise we have introduced constraint matrix $\constraintmat \in \{0,1\}^{|\paramsmap_i|\times\nfg}$, using $|\cdot|$ to denote the length of a vector. We now make concrete our definition of $\paramsmap_I$: they are real-valued vectors of a \emph{shared} length $|\paramsmap_I|$ that, when multiplied by $\constraintmat$, produce a vector of length $\nfg$ that is compatible with existing atomistic constraints. $|\paramsmap_I|$ is less than or equal to $\nfg$ and quantifies the degrees of freedom present after accounting for atomistic constraints. $\constraintmat$ is similar to an identity matrix, except select groups of rows corresponding to atoms participating in constrained bonds have been replaced by their sum along columns. For example, consider a hypothetical system of 4 atoms (i.e., $\nfg=4$), where atoms 1 and 2 (using zero based indexing) participate in a constrained bond: here, $\constraintmat$ is given by $\constraintmat_\mathrm{example}$ in Eq. \eqref{eq:constraintmat_example}. Due to the constrained bond, $|\paramsmap_I|=3$.
\begin{equation}
\constraintmat_\mathrm{example} := 
\begin{pmatrix}
1 & 0 & 0 & 0 \\
0 & 1 & 1 & 0 \\
0 & 0 & 0 & 1 \\
\end{pmatrix}
\label{eq:constraintmat_example}
\end{equation}
 Note that $\constraintmat$ is shared for all CG sites in a given atomistic system, as it is defined only using details of the atomistic system, and that other parameterizations, even those using position dependent features, are possible; we reserve these options for future works.

Together, substitution results in Eq. \eqref{eq:qp_sub_opt_complete},
\begin{equation}
\left[
\sum_{d}
\raverage{
\left[
[\paramsmap_I \constraintmat\outerprod \ident_3]_{i,d} \cdot \aaforce(\rfg)
\right]^2
}
\right]_I
\label{eq:qp_sub_opt_complete}
\end{equation}
which is in turn approximated by a trajectory average in Eq. \eqref{eq:qp_traj_average}, where we have introduced $\trajforcemat$ of size $3n \times n_t$ to contain the atomistic forces present in frames of the trajectory. Note that we have used $\trajforcemat_{:,t}$ to denote extracting matrix columns (in contrast to rows). 
\begin{equation}
\approx
\left[
\frac{1}{n_t}
\sum_t^{n_t}
\sum_{d}
\left[
[\paramsmap_I \constraintmat\outerprod \ident_3]_{d} \cdot \trajforcemat_{:,t}
\right]^2
\right]_I
\label{eq:qp_traj_average}
\end{equation}
For numerical minimization, it is convenient to reshape the involved arrays. This can be performed by introducing a reshaped array containing the atomistic forces of shape $\nfg \times 3 n_t$, denoted $\trajforcematreshape$, which is constructed by stacking the forces along each Cartesian coordinate onto the axis previously used only for indexing trajectory frames. This allows us to remove the outer matrix product as shown in \eqref{eq:qp_reshape}, where we have now made $d$ index over a numerical representation of the Cartesian components.
\begin{equation}
=
\left[
\frac{1}{n_t}
\sum_t^{n_t}
\sum_{d\in\{0,1,2\}}
\left[
\paramsmap_I \constraintmat \cdot \trajforcematreshape_{:,t+d}
\right]^2
\right]_I
\propto
\left[
\|
\paramsmap_I \constraintmat \trajforcematreshape
\|^2_2
\right]_I
\label{eq:qp_reshape}
\end{equation}

While the linear programming constraints related to constrained bonds are satisfied via $\constraintmat$, the final step is to satisfy constraints relative to the mapping operator. As specified in the main text, for linear maps this is expressed as Eq. \eqref{eq:qp_cmap_constraints}.
\begin{equation}
    \forcemapmat \cgmapmat^{T}
    = 
    \ident
    \label{eq:qp_cmap_constraints}
\end{equation}
However, if we assume that the contributions in the configurational map are also particle-wise, i.e. there exists $\pcgmapmat$ such that Eq. \eqref{eq:pcmap} holds, we may define our constraints more concisely. The particle-wise form of this, completed with substitution from Eq. \eqref{eq:qp_force_map_param}, is shown in Eq. \eqref{eq:qp_cmap_constraints_pwise}, where $e_I$ denotes a one-hot vector. 

\begin{equation}
    \pcgmapmat\outerprod \ident_3 = \cgmapmat
    \label{eq:pcmap}
\end{equation}

\begin{equation}
    \paramsmap_I \constraintmat \pcgmapmat^{T} 
    = 
    e_I
    \label{eq:qp_cmap_constraints_pwise}
\end{equation}
After optimization, each row of the particle-wise force map may be reconstructed via Eq. \eqref{eq:qp_force_map_param}, and the full force map may be obtained by using Eq. \eqref{eq:pforcemapmat}.

It is important to note that the above equations directly optimize the residue as a sample average over a trajectory instead of as the population average described by Eq. \eqref{eq:linear_force_residual}. The relationship between the approximated and true residuals is captured by the concept of overfitting in ML, and we similarly use holdout sets to ensure that our optimised force maps improve the corresponding population values. However, as the number of free parameters in the proposed force optimization examples (hundreds) is significantly less than the number of trajectory (millions), we observe no practical difference in hold-out and train force residuals upon force optimization.


\begin{mcitethebibliography}{92}
\providecommand*\natexlab[1]{#1}
\providecommand*\mciteSetBstSublistMode[1]{}
\providecommand*\mciteSetBstMaxWidthForm[2]{}
\providecommand*\mciteBstWouldAddEndPuncttrue
  {\def\EndOfBibitem{\unskip.}}
\providecommand*\mciteBstWouldAddEndPunctfalse
  {\let\EndOfBibitem\relax}
\providecommand*\mciteSetBstMidEndSepPunct[3]{}
\providecommand*\mciteSetBstSublistLabelBeginEnd[3]{}
\providecommand*\EndOfBibitem{}
\mciteSetBstSublistMode{f}
\mciteSetBstMaxWidthForm{subitem}{(\alph{mcitesubitemcount})}
\mciteSetBstSublistLabelBeginEnd
  {\mcitemaxwidthsubitemform\space}
  {\relax}
  {\relax}

\bibitem[Hollingsworth and Dror(2018)Hollingsworth, and
  Dror]{hollingsworth2018molecular}
Hollingsworth,~S.~A.; Dror,~R.~O. Molecular dynamics simulation for all.
  \emph{Neuron} \textbf{2018}, \emph{99}, 1129--1143\relax
\mciteBstWouldAddEndPuncttrue
\mciteSetBstMidEndSepPunct{\mcitedefaultmidpunct}
{\mcitedefaultendpunct}{\mcitedefaultseppunct}\relax
\EndOfBibitem
\bibitem[Bottaro and Lindorff-Larsen(2018)Bottaro, and
  Lindorff-Larsen]{bottaro2018biophysical}
Bottaro,~S.; Lindorff-Larsen,~K. Biophysical experiments and biomolecular
  simulations: A perfect match? \emph{Science} \textbf{2018}, \emph{361},
  355--360\relax
\mciteBstWouldAddEndPuncttrue
\mciteSetBstMidEndSepPunct{\mcitedefaultmidpunct}
{\mcitedefaultendpunct}{\mcitedefaultseppunct}\relax
\EndOfBibitem
\bibitem[Gartner~III and Jayaraman(2019)Gartner~III, and
  Jayaraman]{gartner2019modeling}
Gartner~III,~T.~E.; Jayaraman,~A. Modeling and simulations of polymers: a
  roadmap. \emph{Macromolecules} \textbf{2019}, \emph{52}, 755--786\relax
\mciteBstWouldAddEndPuncttrue
\mciteSetBstMidEndSepPunct{\mcitedefaultmidpunct}
{\mcitedefaultendpunct}{\mcitedefaultseppunct}\relax
\EndOfBibitem
\bibitem[Baschnagel \latin{et~al.}(2000)Baschnagel, Binder, Doruker, Gusev,
  Hahn, Kremer, Mattice, M{\"u}ller-Plathe, Murat, Paul, \latin{et~al.}
  others]{baschnagel2000bridging}
Baschnagel,~J.; Binder,~K.; Doruker,~P.; Gusev,~A.~A.; Hahn,~O.; Kremer,~K.;
  Mattice,~W.~L.; M{\"u}ller-Plathe,~F.; Murat,~M.; Paul,~W. \latin{et~al.}
  Bridging the gap between atomistic and coarse-grained models of polymers:
  status and perspectives. \emph{Viscoelasticity, atomistic models, statistical
  chemistry} \textbf{2000}, 41--156\relax
\mciteBstWouldAddEndPuncttrue
\mciteSetBstMidEndSepPunct{\mcitedefaultmidpunct}
{\mcitedefaultendpunct}{\mcitedefaultseppunct}\relax
\EndOfBibitem
\bibitem[Klein and Shinoda(2008)Klein, and Shinoda]{klein2008large}
Klein,~M.~L.; Shinoda,~W. Large-scale molecular dynamics simulations of
  self-assembling systems. \emph{Science} \textbf{2008}, \emph{321},
  798--800\relax
\mciteBstWouldAddEndPuncttrue
\mciteSetBstMidEndSepPunct{\mcitedefaultmidpunct}
{\mcitedefaultendpunct}{\mcitedefaultseppunct}\relax
\EndOfBibitem
\bibitem[Noid(2013)]{noid2013perspective}
Noid,~W.~G. Perspective: Coarse-grained models for biomolecular systems.
  \emph{J. Chem. Phys.} \textbf{2013}, \emph{139}, 09B201\_1\relax
\mciteBstWouldAddEndPuncttrue
\mciteSetBstMidEndSepPunct{\mcitedefaultmidpunct}
{\mcitedefaultendpunct}{\mcitedefaultseppunct}\relax
\EndOfBibitem
\bibitem[Pak and Voth(2018)Pak, and Voth]{pak2018advances}
Pak,~A.~J.; Voth,~G.~A. Advances in coarse-grained modeling of macromolecular
  complexes. \emph{Curr. Opin. Struct. Biol.} \textbf{2018}, \emph{52},
  119--126\relax
\mciteBstWouldAddEndPuncttrue
\mciteSetBstMidEndSepPunct{\mcitedefaultmidpunct}
{\mcitedefaultendpunct}{\mcitedefaultseppunct}\relax
\EndOfBibitem
\bibitem[Dhamankar and Webb(2021)Dhamankar, and Webb]{dhamankar2021chemically}
Dhamankar,~S.; Webb,~M.~A. Chemically specific coarse-graining of polymers:
  methods and prospects. \emph{J. Polym. Sci.} \textbf{2021}, \emph{59},
  2613--2643\relax
\mciteBstWouldAddEndPuncttrue
\mciteSetBstMidEndSepPunct{\mcitedefaultmidpunct}
{\mcitedefaultendpunct}{\mcitedefaultseppunct}\relax
\EndOfBibitem
\bibitem[Jin \latin{et~al.}(2022)Jin, Pak, Durumeric, Loose, and
  Voth]{jin2022bottom}
Jin,~J.; Pak,~A.~J.; Durumeric,~A.~E.; Loose,~T.~D.; Voth,~G.~A. Bottom-up
  Coarse-Graining: Principles and Perspectives. \emph{J. Chem. Theory Comput.}
  \textbf{2022}, \emph{18}, 5759--5791\relax
\mciteBstWouldAddEndPuncttrue
\mciteSetBstMidEndSepPunct{\mcitedefaultmidpunct}
{\mcitedefaultendpunct}{\mcitedefaultseppunct}\relax
\EndOfBibitem
\bibitem[Lemke and Peter(2017)Lemke, and Peter]{lemke2017neural}
Lemke,~T.; Peter,~C. Neural network based prediction of conformational free
  energies-a new route toward coarse-grained simulation models. \emph{J. Chem.
  Theory Comput.} \textbf{2017}, \emph{13}, 6213--6221\relax
\mciteBstWouldAddEndPuncttrue
\mciteSetBstMidEndSepPunct{\mcitedefaultmidpunct}
{\mcitedefaultendpunct}{\mcitedefaultseppunct}\relax
\EndOfBibitem
\bibitem[Zhang \latin{et~al.}(2018)Zhang, Han, Wang, Car, and
  Weinan]{zhang2018deepcg}
Zhang,~L.; Han,~J.; Wang,~H.; Car,~R.; Weinan,~W.~E. DeePCG: Constructing
  coarse-grained models via deep neural networks. \emph{J. Chem. Phys.}
  \textbf{2018}, \emph{149}, 034101\relax
\mciteBstWouldAddEndPuncttrue
\mciteSetBstMidEndSepPunct{\mcitedefaultmidpunct}
{\mcitedefaultendpunct}{\mcitedefaultseppunct}\relax
\EndOfBibitem
\bibitem[Wang \latin{et~al.}(2019)Wang, Olsson, Wehmeyer, Pérez, Charron,
  Fabritiis, Noé, and Clementi]{Wang2019}
Wang,~J.; Olsson,~S.; Wehmeyer,~C.; Pérez,~A.; Charron,~N.~E.;
  Fabritiis,~G.~D.; Noé,~F.; Clementi,~C. Machine Learning of Coarse-Grained
  Molecular Dynamics Force Fields. \emph{ACS Cent. Sci.} \textbf{2019},
  \emph{5}, 755--767\relax
\mciteBstWouldAddEndPuncttrue
\mciteSetBstMidEndSepPunct{\mcitedefaultmidpunct}
{\mcitedefaultendpunct}{\mcitedefaultseppunct}\relax
\EndOfBibitem
\bibitem[Wang and G{\'o}mez-Bombarelli(2019)Wang, and
  G{\'o}mez-Bombarelli]{wang2019cgautoencoders}
Wang,~W.; G{\'o}mez-Bombarelli,~R. Coarse-graining auto-encoders for molecular
  dynamics. \emph{npj Comput. Mater.} \textbf{2019}, \emph{5}, 125\relax
\mciteBstWouldAddEndPuncttrue
\mciteSetBstMidEndSepPunct{\mcitedefaultmidpunct}
{\mcitedefaultendpunct}{\mcitedefaultseppunct}\relax
\EndOfBibitem
\bibitem[Husic \latin{et~al.}(2020)Husic, Charron, Lemm, Wang, Pérez,
  Majewski, Krämer, Chen, Olsson, Fabritiis, Noé, and Clementi]{Husic2020}
Husic,~B.~E.; Charron,~N.~E.; Lemm,~D.; Wang,~J.; Pérez,~A.; Majewski,~M.;
  Krämer,~A.; Chen,~Y.; Olsson,~S.; Fabritiis,~G.~D. \latin{et~al.}  Coarse
  graining molecular dynamics with graph neural networks. \emph{J. Chem. Phys.}
  \textbf{2020}, \emph{153}\relax
\mciteBstWouldAddEndPuncttrue
\mciteSetBstMidEndSepPunct{\mcitedefaultmidpunct}
{\mcitedefaultendpunct}{\mcitedefaultseppunct}\relax
\EndOfBibitem
\bibitem[Wang \latin{et~al.}(2021)Wang, Charron, Husic, Olsson, Noé, and
  Clementi]{wang2021multibody}
Wang,~J.; Charron,~N.; Husic,~B.; Olsson,~S.; Noé,~F.; Clementi,~C. Multi-body
  effects in a coarse-grained protein force field. \emph{J. Chem. Phys.}
  \textbf{2021}, \emph{154}\relax
\mciteBstWouldAddEndPuncttrue
\mciteSetBstMidEndSepPunct{\mcitedefaultmidpunct}
{\mcitedefaultendpunct}{\mcitedefaultseppunct}\relax
\EndOfBibitem
\bibitem[Chen \latin{et~al.}(2021)Chen, Krämer, Charron, Husic, Clementi, and
  Noé]{Chen2021implicit}
Chen,~Y.; Krämer,~A.; Charron,~N.~E.; Husic,~B.~E.; Clementi,~C.; Noé,~F.
  Machine learning implicit solvation for molecular dynamics. \emph{J. Chem.
  Phys.} \textbf{2021}, \emph{155}, 084101\relax
\mciteBstWouldAddEndPuncttrue
\mciteSetBstMidEndSepPunct{\mcitedefaultmidpunct}
{\mcitedefaultendpunct}{\mcitedefaultseppunct}\relax
\EndOfBibitem
\bibitem[Chennakesavalu \latin{et~al.}(2022)Chennakesavalu, Toomer, and
  Rotskoff]{Chennakesavalu2022ensuring}
Chennakesavalu,~S.; Toomer,~D.~J.; Rotskoff,~G.~M. Ensuring thermodynamic
  consistency with invertible coarse-graining. \emph{arXiv preprint
  arXiv:2210.07882} \textbf{2022}, \relax
\mciteBstWouldAddEndPunctfalse
\mciteSetBstMidEndSepPunct{\mcitedefaultmidpunct}
{}{\mcitedefaultseppunct}\relax
\EndOfBibitem
\bibitem[Majewski \latin{et~al.}(2022)Majewski, P{\'e}rez, Th{\"o}lke, Doerr,
  Charron, Giorgino, Husic, Clementi, No{\'e}, and
  De~Fabritiis]{majewski2022machine}
Majewski,~M.; P{\'e}rez,~A.; Th{\"o}lke,~P.; Doerr,~S.; Charron,~N.~E.;
  Giorgino,~T.; Husic,~B.~E.; Clementi,~C.; No{\'e},~F.; De~Fabritiis,~G.
  Machine Learning Coarse-Grained Potentials of Protein Thermodynamics.
  \emph{arXiv preprint arXiv:2212.07492} \textbf{2022}, \relax
\mciteBstWouldAddEndPunctfalse
\mciteSetBstMidEndSepPunct{\mcitedefaultmidpunct}
{}{\mcitedefaultseppunct}\relax
\EndOfBibitem
\bibitem[Ding and Zhang(2022)Ding, and Zhang]{Ding2022coarsegrained}
Ding,~X.; Zhang,~B. Contrastive Learning of Coarse-Grained Force Fields.
  \emph{J. Chem. Theory Comput.} \textbf{2022}, \emph{18}, 6334--6344\relax
\mciteBstWouldAddEndPuncttrue
\mciteSetBstMidEndSepPunct{\mcitedefaultmidpunct}
{\mcitedefaultendpunct}{\mcitedefaultseppunct}\relax
\EndOfBibitem
\bibitem[Durumeric \latin{et~al.}(2023)Durumeric, Charron, Templeton, Musil,
  Bonneau, Pasos-Trejo, Chen, Kelkar, No{\'e}, and
  Clementi]{durumeric2023machine}
Durumeric,~A.~E.; Charron,~N.~E.; Templeton,~C.; Musil,~F.; Bonneau,~K.;
  Pasos-Trejo,~A.~S.; Chen,~Y.; Kelkar,~A.; No{\'e},~F.; Clementi,~C. Machine
  learned coarse-grained protein force-fields: Are we there yet? \emph{Curr.
  Opin. Struct. Biol.} \textbf{2023}, \emph{79}, 102533\relax
\mciteBstWouldAddEndPuncttrue
\mciteSetBstMidEndSepPunct{\mcitedefaultmidpunct}
{\mcitedefaultendpunct}{\mcitedefaultseppunct}\relax
\EndOfBibitem
\bibitem[Yao \latin{et~al.}(2023)Yao, Van, Pan, Park, Mao, Pu, Mei, and
  Shao]{yao2023machine}
Yao,~S.; Van,~R.; Pan,~X.; Park,~J.~H.; Mao,~Y.; Pu,~J.; Mei,~Y.; Shao,~Y.
  Machine learning based implicit solvent model for aqueous-solution alanine
  dipeptide molecular dynamics simulations. \emph{RSC Adv.} \textbf{2023},
  \emph{13}, 4565--4577\relax
\mciteBstWouldAddEndPuncttrue
\mciteSetBstMidEndSepPunct{\mcitedefaultmidpunct}
{\mcitedefaultendpunct}{\mcitedefaultseppunct}\relax
\EndOfBibitem
\bibitem[Joshi and Deshmukh(2021)Joshi, and Deshmukh]{joshi2021review}
Joshi,~S.~Y.; Deshmukh,~S.~A. A review of advancements in coarse-grained
  molecular dynamics simulations. \emph{Mol. Simul.} \textbf{2021}, \emph{47},
  786--803\relax
\mciteBstWouldAddEndPuncttrue
\mciteSetBstMidEndSepPunct{\mcitedefaultmidpunct}
{\mcitedefaultendpunct}{\mcitedefaultseppunct}\relax
\EndOfBibitem
\bibitem[Schommers(1973)]{schommers1973pair}
Schommers,~W. A pair potential for liquid rubidium from the pair correlation
  function. \emph{Phys. Lett. A} \textbf{1973}, \emph{43}, 157--158\relax
\mciteBstWouldAddEndPuncttrue
\mciteSetBstMidEndSepPunct{\mcitedefaultmidpunct}
{\mcitedefaultendpunct}{\mcitedefaultseppunct}\relax
\EndOfBibitem
\bibitem[Lyubartsev and Laaksonen(1995)Lyubartsev, and
  Laaksonen]{lyubartsev1995calculation}
Lyubartsev,~A.~P.; Laaksonen,~A. Calculation of effective interaction
  potentials from radial distribution functions: A reverse Monte Carlo
  approach. \emph{Phys. Rev. E} \textbf{1995}, \emph{52}, 3730\relax
\mciteBstWouldAddEndPuncttrue
\mciteSetBstMidEndSepPunct{\mcitedefaultmidpunct}
{\mcitedefaultendpunct}{\mcitedefaultseppunct}\relax
\EndOfBibitem
\bibitem[M{\"u}ller-Plathe(2002)]{muller2002coarse}
M{\"u}ller-Plathe,~F. Coarse-graining in polymer simulation: from the atomistic
  to the mesoscopic scale and back. \emph{ChemPhysChem} \textbf{2002},
  \emph{3}, 754--769\relax
\mciteBstWouldAddEndPuncttrue
\mciteSetBstMidEndSepPunct{\mcitedefaultmidpunct}
{\mcitedefaultendpunct}{\mcitedefaultseppunct}\relax
\EndOfBibitem
\bibitem[T{\'o}th(2007)]{toth2007interactions}
T{\'o}th,~G. Interactions from diffraction data: historical and comprehensive
  overview of simulation assisted methods. \emph{J. Phys. Condens. Matter}
  \textbf{2007}, \emph{19}, 335220\relax
\mciteBstWouldAddEndPuncttrue
\mciteSetBstMidEndSepPunct{\mcitedefaultmidpunct}
{\mcitedefaultendpunct}{\mcitedefaultseppunct}\relax
\EndOfBibitem
\bibitem[Shell(2008)]{shell2008relative}
Shell,~M.~S. The relative entropy is fundamental to multiscale and inverse
  thermodynamic problems. \emph{J. Chem. Phys.} \textbf{2008}, \emph{129},
  144108\relax
\mciteBstWouldAddEndPuncttrue
\mciteSetBstMidEndSepPunct{\mcitedefaultmidpunct}
{\mcitedefaultendpunct}{\mcitedefaultseppunct}\relax
\EndOfBibitem
\bibitem[Cho and Chu(2009)Cho, and Chu]{cho2009inversion}
Cho,~H.~M.; Chu,~J.-W. Inversion of radial distribution functions to pair
  forces by solving the Yvon--Born--Green equation iteratively. \emph{J. Chem.
  Phys.} \textbf{2009}, \emph{131}, 134107\relax
\mciteBstWouldAddEndPuncttrue
\mciteSetBstMidEndSepPunct{\mcitedefaultmidpunct}
{\mcitedefaultendpunct}{\mcitedefaultseppunct}\relax
\EndOfBibitem
\bibitem[Lu \latin{et~al.}(2013)Lu, Dama, and Voth]{lu2013fitting}
Lu,~L.; Dama,~J.~F.; Voth,~G.~A. Fitting coarse-grained distribution functions
  through an iterative force-matching method. \emph{J. Chem. Phys.}
  \textbf{2013}, \emph{139}, 09B606\_1\relax
\mciteBstWouldAddEndPuncttrue
\mciteSetBstMidEndSepPunct{\mcitedefaultmidpunct}
{\mcitedefaultendpunct}{\mcitedefaultseppunct}\relax
\EndOfBibitem
\bibitem[Rudzinski and Noid(2014)Rudzinski, and
  Noid]{rudzinski2014investigation}
Rudzinski,~J.~F.; Noid,~W.~G. Investigation of coarse-grained mappings via an
  iterative generalized Yvon-Born-Green method. \emph{J. Phys. Chem. B}
  \textbf{2014}, \emph{118}, 8295--8312\relax
\mciteBstWouldAddEndPuncttrue
\mciteSetBstMidEndSepPunct{\mcitedefaultmidpunct}
{\mcitedefaultendpunct}{\mcitedefaultseppunct}\relax
\EndOfBibitem
\bibitem[Sch{\"o}berl \latin{et~al.}(2017)Sch{\"o}berl, Zabaras, and
  Koutsourelakis]{schoberl2017predictive}
Sch{\"o}berl,~M.; Zabaras,~N.; Koutsourelakis,~P.-S. Predictive
  coarse-graining. \emph{J. Comput. Phys.} \textbf{2017}, \emph{333},
  49--77\relax
\mciteBstWouldAddEndPuncttrue
\mciteSetBstMidEndSepPunct{\mcitedefaultmidpunct}
{\mcitedefaultendpunct}{\mcitedefaultseppunct}\relax
\EndOfBibitem
\bibitem[Thaler and Zavadlav(2021)Thaler, and Zavadlav]{thaler2021learning}
Thaler,~S.; Zavadlav,~J. Learning neural network potentials from experimental
  data via Differentiable Trajectory Reweighting. \emph{Nat. Commun.}
  \textbf{2021}, \emph{12}, 6884\relax
\mciteBstWouldAddEndPuncttrue
\mciteSetBstMidEndSepPunct{\mcitedefaultmidpunct}
{\mcitedefaultendpunct}{\mcitedefaultseppunct}\relax
\EndOfBibitem
\bibitem[Thaler \latin{et~al.}(2022)Thaler, Stupp, and
  Zavadlav]{Thaler2022deep}
Thaler,~S.; Stupp,~M.; Zavadlav,~J. Deep coarse-grained potentials via relative
  entropy minimization. \emph{J. Chem. Phys.} \textbf{2022}, \emph{157},
  244103\relax
\mciteBstWouldAddEndPuncttrue
\mciteSetBstMidEndSepPunct{\mcitedefaultmidpunct}
{\mcitedefaultendpunct}{\mcitedefaultseppunct}\relax
\EndOfBibitem
\bibitem[Izvekov and Voth(2005)Izvekov, and Voth]{izvekov2005multiscale}
Izvekov,~S.; Voth,~G.~A. A multiscale coarse-graining method for biomolecular
  systems. \emph{J. Phys. Chem. B} \textbf{2005}, \emph{109}, 2469--2473\relax
\mciteBstWouldAddEndPuncttrue
\mciteSetBstMidEndSepPunct{\mcitedefaultmidpunct}
{\mcitedefaultendpunct}{\mcitedefaultseppunct}\relax
\EndOfBibitem
\bibitem[Noid \latin{et~al.}(2008)Noid, Chu, Ayton, Krishna, Izvekov, Voth,
  Das, and Andersen]{Noid2008}
Noid,~W.~G.; Chu,~J.~W.; Ayton,~G.~S.; Krishna,~V.; Izvekov,~S.; Voth,~G.~A.;
  Das,~A.; Andersen,~H.~C. The multiscale coarse-graining method. I. A rigorous
  bridge between atomistic and coarse-grained models. \emph{J. Chem. Phys.}
  \textbf{2008}, \emph{128}, 244114\relax
\mciteBstWouldAddEndPuncttrue
\mciteSetBstMidEndSepPunct{\mcitedefaultmidpunct}
{\mcitedefaultendpunct}{\mcitedefaultseppunct}\relax
\EndOfBibitem
\bibitem[Lu and Voth(2012)Lu, and Voth]{lu2012multiscale}
Lu,~L.; Voth,~G.~A. The Multiscale Coarse-Graining Method. \emph{Adv. Chem.
  Phys.} \textbf{2012}, \emph{149}, 47--81\relax
\mciteBstWouldAddEndPuncttrue
\mciteSetBstMidEndSepPunct{\mcitedefaultmidpunct}
{\mcitedefaultendpunct}{\mcitedefaultseppunct}\relax
\EndOfBibitem
\bibitem[Zhang \latin{et~al.}(2008)Zhang, Lu, Noid, Krishna, Pfaendtner, and
  Voth]{zhang2008systematic}
Zhang,~Z.; Lu,~L.; Noid,~W.~G.; Krishna,~V.; Pfaendtner,~J.; Voth,~G.~A. A
  systematic methodology for defining coarse-grained sites in large
  biomolecules. \emph{Biophys. J.} \textbf{2008}, \emph{95}, 5073--5083\relax
\mciteBstWouldAddEndPuncttrue
\mciteSetBstMidEndSepPunct{\mcitedefaultmidpunct}
{\mcitedefaultendpunct}{\mcitedefaultseppunct}\relax
\EndOfBibitem
\bibitem[Cao and Voth(2015)Cao, and Voth]{Cao2015mscgXI}
Cao,~Z.; Voth,~G.~A. The multiscale coarse-graining method. XI. Accurate
  interactions based on the centers of charge of coarse-grained sites. \emph{J.
  Chem. Phys.} \textbf{2015}, \emph{143}, 243116\relax
\mciteBstWouldAddEndPuncttrue
\mciteSetBstMidEndSepPunct{\mcitedefaultmidpunct}
{\mcitedefaultendpunct}{\mcitedefaultseppunct}\relax
\EndOfBibitem
\bibitem[Foley \latin{et~al.}(2015)Foley, Shell, and Noid]{Foley2015resolution}
Foley,~T.~T.; Shell,~M.~S.; Noid,~W.~G. The impact of resolution upon entropy
  and information in coarse-grained models. \emph{J. Chem. Phys.}
  \textbf{2015}, \emph{143}, 243104\relax
\mciteBstWouldAddEndPuncttrue
\mciteSetBstMidEndSepPunct{\mcitedefaultmidpunct}
{\mcitedefaultendpunct}{\mcitedefaultseppunct}\relax
\EndOfBibitem
\bibitem[Madsen \latin{et~al.}(2017)Madsen, Sinitskiy, Li, and
  Voth]{madsen2017highly}
Madsen,~J.~J.; Sinitskiy,~A.~V.; Li,~J.; Voth,~G.~A. Highly coarse-grained
  representations of transmembrane proteins. \emph{J. Chem. Theory Comput.}
  \textbf{2017}, \emph{13}, 935--944\relax
\mciteBstWouldAddEndPuncttrue
\mciteSetBstMidEndSepPunct{\mcitedefaultmidpunct}
{\mcitedefaultendpunct}{\mcitedefaultseppunct}\relax
\EndOfBibitem
\bibitem[Diggins~IV \latin{et~al.}(2018)Diggins~IV, Liu, Deserno, and
  Potestio]{diggins2018optimal}
Diggins~IV,~P.; Liu,~C.; Deserno,~M.; Potestio,~R. Optimal coarse-grained site
  selection in elastic network models of biomolecules. \emph{J. Chem. Theory
  Comput.} \textbf{2018}, \emph{15}, 648--664\relax
\mciteBstWouldAddEndPuncttrue
\mciteSetBstMidEndSepPunct{\mcitedefaultmidpunct}
{\mcitedefaultendpunct}{\mcitedefaultseppunct}\relax
\EndOfBibitem
\bibitem[Webb \latin{et~al.}(2019)Webb, Delannoy, and
  de~Pablo]{webb2019graphbased}
Webb,~M.~A.; Delannoy,~J.-Y.; de~Pablo,~J.~J. Graph-Based Approach to
  Systematic Molecular Coarse-Graining. \emph{J. Chem. Theory Comput.}
  \textbf{2019}, \emph{15}, 1199--1208\relax
\mciteBstWouldAddEndPuncttrue
\mciteSetBstMidEndSepPunct{\mcitedefaultmidpunct}
{\mcitedefaultendpunct}{\mcitedefaultseppunct}\relax
\EndOfBibitem
\bibitem[Giulini \latin{et~al.}(2020)Giulini, Menichetti, Shell, and
  Potestio]{giulini2020informationtheorybased}
Giulini,~M.; Menichetti,~R.; Shell,~M.~S.; Potestio,~R. An
  Information-Theory-Based Approach for Optimal Model Reduction of
  Biomolecules. \emph{J. Chem. Theory Comput.} \textbf{2020}, \emph{16},
  6795--6813\relax
\mciteBstWouldAddEndPuncttrue
\mciteSetBstMidEndSepPunct{\mcitedefaultmidpunct}
{\mcitedefaultendpunct}{\mcitedefaultseppunct}\relax
\EndOfBibitem
\bibitem[Souza \latin{et~al.}(2021)Souza, Alessandri, Barnoud, Thallmair,
  Faustino, Gr{\"u}newald, Patmanidis, Abdizadeh, Bruininks, Wassenaar,
  \latin{et~al.} others]{souza2021martini3}
Souza,~P.~C.; Alessandri,~R.; Barnoud,~J.; Thallmair,~S.; Faustino,~I.;
  Gr{\"u}newald,~F.; Patmanidis,~I.; Abdizadeh,~H.; Bruininks,~B.~M.;
  Wassenaar,~T.~A. \latin{et~al.}  Martini 3: a general purpose force field for
  coarse-grained molecular dynamics. \emph{Nat. Methods} \textbf{2021},
  \emph{18}, 382--388\relax
\mciteBstWouldAddEndPuncttrue
\mciteSetBstMidEndSepPunct{\mcitedefaultmidpunct}
{\mcitedefaultendpunct}{\mcitedefaultseppunct}\relax
\EndOfBibitem
\bibitem[Kidder \latin{et~al.}(2021)Kidder, Szukalo, and
  Noid]{kidder2021energetic}
Kidder,~K.~M.; Szukalo,~R.~J.; Noid,~W. Energetic and entropic considerations
  for coarse-graining. \emph{Eur. Phys. J. B} \textbf{2021}, \emph{94},
  153\relax
\mciteBstWouldAddEndPuncttrue
\mciteSetBstMidEndSepPunct{\mcitedefaultmidpunct}
{\mcitedefaultendpunct}{\mcitedefaultseppunct}\relax
\EndOfBibitem
\bibitem[Yang \latin{et~al.}(2023)Yang, Templeton, Rosenberger, Bittracher,
  Nüske, Noé, and Clementi]{yang2023slicing}
Yang,~W.; Templeton,~C.; Rosenberger,~D.; Bittracher,~A.; Nüske,~F.; Noé,~F.;
  Clementi,~C. Slicing and Dicing: Optimal Coarse-Grained Representation to
  Preserve Molecular Kinetics. \emph{ACS Cent. Sci.} \textbf{2023}, \relax
\mciteBstWouldAddEndPunctfalse
\mciteSetBstMidEndSepPunct{\mcitedefaultmidpunct}
{}{\mcitedefaultseppunct}\relax
\EndOfBibitem
\bibitem[Larini \latin{et~al.}(2010)Larini, Lu, and Voth]{larini2010multiscale}
Larini,~L.; Lu,~L.; Voth,~G.~A. The multiscale coarse-graining method. VI.
  Implementation of three-body coarse-grained potentials. \emph{J. Chem. Phys.}
  \textbf{2010}, \emph{132}, 164107\relax
\mciteBstWouldAddEndPuncttrue
\mciteSetBstMidEndSepPunct{\mcitedefaultmidpunct}
{\mcitedefaultendpunct}{\mcitedefaultseppunct}\relax
\EndOfBibitem
\bibitem[Sanyal and Shell(2016)Sanyal, and Shell]{sanyal2016coarse}
Sanyal,~T.; Shell,~M.~S. Coarse-grained models using local-density potentials
  optimized with the relative entropy: Application to implicit solvation.
  \emph{J. Chem. Phys.} \textbf{2016}, \emph{145}, 034109\relax
\mciteBstWouldAddEndPuncttrue
\mciteSetBstMidEndSepPunct{\mcitedefaultmidpunct}
{\mcitedefaultendpunct}{\mcitedefaultseppunct}\relax
\EndOfBibitem
\bibitem[John and Cs{\'a}nyi(2017)John, and Cs{\'a}nyi]{john2017many}
John,~S.; Cs{\'a}nyi,~G. Many-body coarse-grained interactions using Gaussian
  approximation potentials. \emph{J. Phys. Chem. B} \textbf{2017}, \emph{121},
  10934--10949\relax
\mciteBstWouldAddEndPuncttrue
\mciteSetBstMidEndSepPunct{\mcitedefaultmidpunct}
{\mcitedefaultendpunct}{\mcitedefaultseppunct}\relax
\EndOfBibitem
\bibitem[Scherer and Andrienko(2018)Scherer, and
  Andrienko]{scherer2018understanding}
Scherer,~C.; Andrienko,~D. Understanding three-body contributions to
  coarse-grained force fields. \emph{Phys. Chem. Chem. Phys.} \textbf{2018},
  \emph{20}, 22387--22394\relax
\mciteBstWouldAddEndPuncttrue
\mciteSetBstMidEndSepPunct{\mcitedefaultmidpunct}
{\mcitedefaultendpunct}{\mcitedefaultseppunct}\relax
\EndOfBibitem
\bibitem[DeLyser and Noid(2022)DeLyser, and Noid]{delyser2022coarse}
DeLyser,~M.~R.; Noid,~W. Coarse-grained models for local density gradients.
  \emph{J. Chem. Phys.} \textbf{2022}, \emph{156}, 034106\relax
\mciteBstWouldAddEndPuncttrue
\mciteSetBstMidEndSepPunct{\mcitedefaultmidpunct}
{\mcitedefaultendpunct}{\mcitedefaultseppunct}\relax
\EndOfBibitem
\bibitem[Dama \latin{et~al.}(2013)Dama, Sinitskiy, McCullagh, Weare, Roux,
  Dinner, and Voth]{dama2013theory}
Dama,~J.~F.; Sinitskiy,~A.~V.; McCullagh,~M.; Weare,~J.; Roux,~B.;
  Dinner,~A.~R.; Voth,~G.~A. The theory of ultra-coarse-graining. 1. General
  principles. \emph{J. Chem. Theory Comput.} \textbf{2013}, \emph{9},
  2466--2480\relax
\mciteBstWouldAddEndPuncttrue
\mciteSetBstMidEndSepPunct{\mcitedefaultmidpunct}
{\mcitedefaultendpunct}{\mcitedefaultseppunct}\relax
\EndOfBibitem
\bibitem[Sharp \latin{et~al.}(2019)Sharp, V{\'a}zquez, Wagner,
  Dannenhoffer-Lafage, and Voth]{sharp2019multiconfigurational}
Sharp,~M.~E.; V{\'a}zquez,~F.~X.; Wagner,~J.~W.; Dannenhoffer-Lafage,~T.;
  Voth,~G.~A. Multiconfigurational coarse-grained molecular dynamics. \emph{J.
  Chem. Theory Comput.} \textbf{2019}, \emph{15}, 3306--3315\relax
\mciteBstWouldAddEndPuncttrue
\mciteSetBstMidEndSepPunct{\mcitedefaultmidpunct}
{\mcitedefaultendpunct}{\mcitedefaultseppunct}\relax
\EndOfBibitem
\bibitem[Rudzinski and Bereau(2020)Rudzinski, and Bereau]{rudzinski2020coarse}
Rudzinski,~J.~F.; Bereau,~T. Coarse-grained conformational surface hopping:
  Methodology and transferability. \emph{J. Chem. Phys.} \textbf{2020},
  \emph{153}, 214110\relax
\mciteBstWouldAddEndPuncttrue
\mciteSetBstMidEndSepPunct{\mcitedefaultmidpunct}
{\mcitedefaultendpunct}{\mcitedefaultseppunct}\relax
\EndOfBibitem
\bibitem[Jin \latin{et~al.}(2021)Jin, Han, Pak, and Voth]{jin2021new}
Jin,~J.; Han,~Y.; Pak,~A.~J.; Voth,~G.~A. A new one-site coarse-grained model
  for water: Bottom-up many-body projected water (BUMPer). I. General theory
  and model. \emph{J. Chem. Phys.} \textbf{2021}, \emph{154}, 044104\relax
\mciteBstWouldAddEndPuncttrue
\mciteSetBstMidEndSepPunct{\mcitedefaultmidpunct}
{\mcitedefaultendpunct}{\mcitedefaultseppunct}\relax
\EndOfBibitem
\bibitem[Sahrmann \latin{et~al.}(2022)Sahrmann, Loose, Durumeric, and
  Voth]{sahrmann2022utilizing}
Sahrmann,~P.~G.; Loose,~T.~D.; Durumeric,~A.~E.; Voth,~G.~A. Utilizing Machine
  Learning to Greatly Expand the Range and Accuracy of Bottom-Up Coarse-Grained
  Models Through Virtual Particles. \emph{arXiv preprint arXiv:2212.04530}
  \textbf{2022}, \relax
\mciteBstWouldAddEndPunctfalse
\mciteSetBstMidEndSepPunct{\mcitedefaultmidpunct}
{}{\mcitedefaultseppunct}\relax
\EndOfBibitem
\bibitem[Mohri \latin{et~al.}(2018)Mohri, Rostamizadeh, and
  Talwalkar]{mohri2018foundations}
Mohri,~M.; Rostamizadeh,~A.; Talwalkar,~A. \emph{Foundations of machine
  learning}; MIT press, 2018\relax
\mciteBstWouldAddEndPuncttrue
\mciteSetBstMidEndSepPunct{\mcitedefaultmidpunct}
{\mcitedefaultendpunct}{\mcitedefaultseppunct}\relax
\EndOfBibitem
\bibitem[Ciccotti \latin{et~al.}(2005)Ciccotti, Kapral, and
  Vanden-Eijnden]{Ciccotti2005}
Ciccotti,~G.; Kapral,~R.; Vanden-Eijnden,~E. Blue Moon sampling, vectorial
  reaction coordinates, and unbiased constrained dynamics. \emph{ChemPhysChem}
  \textbf{2005}, \emph{6}, 1809--1814, Expression of potential of mean force,
  including under constraints\relax
\mciteBstWouldAddEndPuncttrue
\mciteSetBstMidEndSepPunct{\mcitedefaultmidpunct}
{\mcitedefaultendpunct}{\mcitedefaultseppunct}\relax
\EndOfBibitem
\bibitem[Köhler \latin{et~al.}(2023)Köhler, Chen, Krämer, Clementi, and
  Noé]{flowmatching2023}
Köhler,~J.; Chen,~Y.; Krämer,~A.; Clementi,~C.; Noé,~F. Flow-Matching:
  Efficient Coarse-Graining of Molecular Dynamics without Forces. \emph{J.
  Chem. Theory Comput.} \textbf{2023}, \relax
\mciteBstWouldAddEndPunctfalse
\mciteSetBstMidEndSepPunct{\mcitedefaultmidpunct}
{}{\mcitedefaultseppunct}\relax
\EndOfBibitem
\bibitem[Liu \latin{et~al.}(2008)Liu, Shi, Daum{\'e}~III, and
  Voth]{liu2008bayesian}
Liu,~P.; Shi,~Q.; Daum{\'e}~III,~H.; Voth,~G.~A. A Bayesian statistics approach
  to multiscale coarse graining. \emph{J. Chem. Phys.} \textbf{2008},
  \emph{129}, 12B605\relax
\mciteBstWouldAddEndPuncttrue
\mciteSetBstMidEndSepPunct{\mcitedefaultmidpunct}
{\mcitedefaultendpunct}{\mcitedefaultseppunct}\relax
\EndOfBibitem
\bibitem[Lu \latin{et~al.}(2010)Lu, Izvekov, Das, Andersen, and
  Voth]{lu2010efficient}
Lu,~L.; Izvekov,~S.; Das,~A.; Andersen,~H.~C.; Voth,~G.~A. Efficient,
  regularized, and scalable algorithms for multiscale coarse-graining. \emph{J.
  Chem. Theory Comput.} \textbf{2010}, \emph{6}, 954--965\relax
\mciteBstWouldAddEndPuncttrue
\mciteSetBstMidEndSepPunct{\mcitedefaultmidpunct}
{\mcitedefaultendpunct}{\mcitedefaultseppunct}\relax
\EndOfBibitem
\bibitem[Eastman \latin{et~al.}(2017)Eastman, Swails, Chodera, McGibbon, Zhao,
  Beauchamp, Wang, Simmonett, Harrigan, Stern, Wiewiora, Brooks, and
  Pande]{eastman2017openmm}
Eastman,~P.; Swails,~J.; Chodera,~J.~D.; McGibbon,~R.~T.; Zhao,~Y.;
  Beauchamp,~K.~A.; Wang,~L.-P.; Simmonett,~A.~C.; Harrigan,~M.~P.;
  Stern,~C.~D. \latin{et~al.}  OpenMM 7: Rapid development of high performance
  algorithms for molecular dynamics. \emph{PLoS Comput. Biol.} \textbf{2017},
  \emph{13}, e1005659\relax
\mciteBstWouldAddEndPuncttrue
\mciteSetBstMidEndSepPunct{\mcitedefaultmidpunct}
{\mcitedefaultendpunct}{\mcitedefaultseppunct}\relax
\EndOfBibitem
\bibitem[Montavon \latin{et~al.}(2012)Montavon, Orr, and
  M{\"u}ller]{montavon2012neural}
Montavon,~G.; Orr,~G.; M{\"u}ller,~K.-R. \emph{Neural networks: tricks of the
  trade}; springer, 2012; Vol. 7700\relax
\mciteBstWouldAddEndPuncttrue
\mciteSetBstMidEndSepPunct{\mcitedefaultmidpunct}
{\mcitedefaultendpunct}{\mcitedefaultseppunct}\relax
\EndOfBibitem
\bibitem[Bottou \latin{et~al.}(2018)Bottou, Curtis, and
  Nocedal]{bottou2018optimization}
Bottou,~L.; Curtis,~F.~E.; Nocedal,~J. Optimization methods for large-scale
  machine learning. \emph{SIAM Rev.} \textbf{2018}, \emph{60}, 223--311\relax
\mciteBstWouldAddEndPuncttrue
\mciteSetBstMidEndSepPunct{\mcitedefaultmidpunct}
{\mcitedefaultendpunct}{\mcitedefaultseppunct}\relax
\EndOfBibitem
\bibitem[Johnson and Zhang(2013)Johnson, and Zhang]{johnson2013svrg}
Johnson,~R.; Zhang,~T. Accelerating stochastic gradient descent using
  predictive variance reduction. \emph{Advances in Neural Information
  Processing Systems} \textbf{2013}, \emph{26}\relax
\mciteBstWouldAddEndPuncttrue
\mciteSetBstMidEndSepPunct{\mcitedefaultmidpunct}
{\mcitedefaultendpunct}{\mcitedefaultseppunct}\relax
\EndOfBibitem
\bibitem[Defazio \latin{et~al.}(2014)Defazio, Bach, and
  Lacoste-Julien]{defazio2014saga}
Defazio,~A.; Bach,~F.; Lacoste-Julien,~S. SAGA: A fast incremental gradient
  method with support for non-strongly convex composite objectives.
  \emph{Advances in Neural Information Processing Systems} \textbf{2014},
  \emph{27}\relax
\mciteBstWouldAddEndPuncttrue
\mciteSetBstMidEndSepPunct{\mcitedefaultmidpunct}
{\mcitedefaultendpunct}{\mcitedefaultseppunct}\relax
\EndOfBibitem
\bibitem[Schmidt \latin{et~al.}(2017)Schmidt, Le~Roux, and
  Bach]{schmidt2017sag}
Schmidt,~M.; Le~Roux,~N.; Bach,~F. Minimizing finite sums with the stochastic
  average gradient. \emph{Math. Program.} \textbf{2017}, \emph{162},
  83--112\relax
\mciteBstWouldAddEndPuncttrue
\mciteSetBstMidEndSepPunct{\mcitedefaultmidpunct}
{\mcitedefaultendpunct}{\mcitedefaultseppunct}\relax
\EndOfBibitem
\bibitem[Nguyen \latin{et~al.}(2017)Nguyen, Liu, Scheinberg, and
  Tak{\'a}{\v{c}}]{nguyen2017sarah}
Nguyen,~L.~M.; Liu,~J.; Scheinberg,~K.; Tak{\'a}{\v{c}},~M. SARAH: A novel
  method for machine learning problems using stochastic recursive gradient.
  International Conference on Machine Learning. 2017; pp 2613--2621\relax
\mciteBstWouldAddEndPuncttrue
\mciteSetBstMidEndSepPunct{\mcitedefaultmidpunct}
{\mcitedefaultendpunct}{\mcitedefaultseppunct}\relax
\EndOfBibitem
\bibitem[Defazio and Bottou(2019)Defazio, and
  Bottou]{defazio2019ineffectiveness}
Defazio,~A.; Bottou,~L. On the ineffectiveness of variance reduced optimization
  for deep learning. \emph{Advances in Neural Information Processing Systems}
  \textbf{2019}, \emph{32}\relax
\mciteBstWouldAddEndPuncttrue
\mciteSetBstMidEndSepPunct{\mcitedefaultmidpunct}
{\mcitedefaultendpunct}{\mcitedefaultseppunct}\relax
\EndOfBibitem
\bibitem[Stellato \latin{et~al.}(2020)Stellato, Banjac, Goulart, Bemporad, and
  Boyd]{osqp}
Stellato,~B.; Banjac,~G.; Goulart,~P.; Bemporad,~A.; Boyd,~S. {OSQP}: an
  operator splitting solver for quadratic programs. \emph{Math. Program.
  Comput.} \textbf{2020}, \emph{12}, 637--672\relax
\mciteBstWouldAddEndPuncttrue
\mciteSetBstMidEndSepPunct{\mcitedefaultmidpunct}
{\mcitedefaultendpunct}{\mcitedefaultseppunct}\relax
\EndOfBibitem
\bibitem[Jorgensen \latin{et~al.}(1983)Jorgensen, Chandrasekhar, Madura, Impey,
  and Klein]{jorgensen1983comparison}
Jorgensen,~W.~L.; Chandrasekhar,~J.; Madura,~J.~D.; Impey,~R.~W.; Klein,~M.~L.
  Comparison of simple potential functions for simulating liquid water.
  \emph{J. Chem. Phys.} \textbf{1983}, \emph{79}, 926--935\relax
\mciteBstWouldAddEndPuncttrue
\mciteSetBstMidEndSepPunct{\mcitedefaultmidpunct}
{\mcitedefaultendpunct}{\mcitedefaultseppunct}\relax
\EndOfBibitem
\bibitem[Naritomi and Fuchigami(2011)Naritomi, and Fuchigami]{Naritomi2011}
Naritomi,~Y.; Fuchigami,~S. Slow dynamics in protein fluctuations revealed by
  time-structure based independent component analysis: The case of domain
  motions. \emph{J. Chem. Phys} \textbf{2011}, \emph{134}, 065101, TICA pioneer
  3/3\relax
\mciteBstWouldAddEndPuncttrue
\mciteSetBstMidEndSepPunct{\mcitedefaultmidpunct}
{\mcitedefaultendpunct}{\mcitedefaultseppunct}\relax
\EndOfBibitem
\bibitem[P{\'e}rez-Hern{\'a}ndez \latin{et~al.}(2013)P{\'e}rez-Hern{\'a}ndez,
  Paul, Giorgino, De~Fabritiis, and No{\'e}]{Perez_JChemPhys2013}
P{\'e}rez-Hern{\'a}ndez,~G.; Paul,~F.; Giorgino,~T.; De~Fabritiis,~G.;
  No{\'e},~F. Identification of slow molecular order parameters for Markov
  model construction. \emph{J. Chem. Phys.} \textbf{2013}, \emph{139},
  07B604\_1\relax
\mciteBstWouldAddEndPuncttrue
\mciteSetBstMidEndSepPunct{\mcitedefaultmidpunct}
{\mcitedefaultendpunct}{\mcitedefaultseppunct}\relax
\EndOfBibitem
\bibitem[Schwantes and Pande(2013)Schwantes, and
  Pande]{Schwantes_JChemTheoryComput2013}
Schwantes,~C.~R.; Pande,~V.~S. Improvements in Markov state model construction
  reveal many non-native interactions in the folding of NTL9. \emph{J. Chem.
  Theory Comput.} \textbf{2013}, \emph{9}, 2000--2009\relax
\mciteBstWouldAddEndPuncttrue
\mciteSetBstMidEndSepPunct{\mcitedefaultmidpunct}
{\mcitedefaultendpunct}{\mcitedefaultseppunct}\relax
\EndOfBibitem
\bibitem[Stocker \latin{et~al.}(2022)Stocker, Gasteiger, Becker, G{\"u}nnemann,
  and Margraf]{stocker2022robust}
Stocker,~S.; Gasteiger,~J.; Becker,~F.; G{\"u}nnemann,~S.; Margraf,~J.~T. How
  robust are modern graph neural network potentials in long and hot molecular
  dynamics simulations? \emph{Mach. Learn.: Sci. Technol.} \textbf{2022},
  \emph{3}, 045010\relax
\mciteBstWouldAddEndPuncttrue
\mciteSetBstMidEndSepPunct{\mcitedefaultmidpunct}
{\mcitedefaultendpunct}{\mcitedefaultseppunct}\relax
\EndOfBibitem
\bibitem[Fu \latin{et~al.}(2022)Fu, Wu, Wang, Xie, Keten, Gomez-Bombarelli, and
  Jaakkola]{fu2022forces}
Fu,~X.; Wu,~Z.; Wang,~W.; Xie,~T.; Keten,~S.; Gomez-Bombarelli,~R.;
  Jaakkola,~T. Forces are not enough: Benchmark and critical evaluation for
  machine learning force fields with molecular simulations. \emph{arXiv
  preprint arXiv:2210.07237} \textbf{2022}, \relax
\mciteBstWouldAddEndPunctfalse
\mciteSetBstMidEndSepPunct{\mcitedefaultmidpunct}
{}{\mcitedefaultseppunct}\relax
\EndOfBibitem
\bibitem[Ricci \latin{et~al.}(2022)Ricci, Giannakopoulos, Karkaletsis,
  Theodorou, and Vergadou]{ricci2022developing}
Ricci,~E.; Giannakopoulos,~G.; Karkaletsis,~V.; Theodorou,~D.~N.; Vergadou,~N.
  Developing Machine-Learned Potentials for Coarse-Grained Molecular
  Simulations: Challenges and Pitfalls. Proceedings of the 12th Hellenic
  Conference on Artificial Intelligence. 2022; pp 1--6\relax
\mciteBstWouldAddEndPuncttrue
\mciteSetBstMidEndSepPunct{\mcitedefaultmidpunct}
{\mcitedefaultendpunct}{\mcitedefaultseppunct}\relax
\EndOfBibitem
\bibitem[Rizzi \latin{et~al.}(2019)Rizzi, Chodera, Naden, Beauchamp, Grinaway,
  Fass, adw62, Rustenburg, Ross, Krämer, Macdonald, Swenson, Simmonett,
  hb0402, and ajsilveira]{rizzi2019openmmtools}
Rizzi,~A.; Chodera,~J.; Naden,~L.; Beauchamp,~K.; Grinaway,~P.; Fass,~J.;
  adw62,; Rustenburg,~B.; Ross,~G.~A.; Krämer,~A. \latin{et~al.}
  {choderalab/openmmtools: 0.19.0}. 2019;
  \url{https://doi.org/10.5281/zenodo.3532826}\relax
\mciteBstWouldAddEndPuncttrue
\mciteSetBstMidEndSepPunct{\mcitedefaultmidpunct}
{\mcitedefaultendpunct}{\mcitedefaultseppunct}\relax
\EndOfBibitem
\bibitem[Paszke \latin{et~al.}(2019)Paszke, Gross, Massa, Lerer, Google,
  Chanan, Killeen, Lin, Gimelshein, Antiga, Desmaison, Xamla, Yang, Devito,
  Nabla, Tejani, Chilamkurthy, Ai, Steiner, Facebook, Facebook, and
  Chintala]{Paszke2019Pytorch}
Paszke,~A.; Gross,~S.; Massa,~F.; Lerer,~A.; Google,~J.~B.; Chanan,~G.;
  Killeen,~T.; Lin,~Z.; Gimelshein,~N.; Antiga,~L. \latin{et~al.}  PyTorch: An
  Imperative Style, High-Performance Deep Learning Library. \emph{Advances in
  Neural Information Processing Systems} \textbf{2019}, \emph{32}\relax
\mciteBstWouldAddEndPuncttrue
\mciteSetBstMidEndSepPunct{\mcitedefaultmidpunct}
{\mcitedefaultendpunct}{\mcitedefaultseppunct}\relax
\EndOfBibitem
\bibitem[Honda \latin{et~al.}(2008)Honda, Akiba, Kato, Sawada, Sekijima,
  Ishimura, Ooishi, Watanabe, Odahara, and Harata]{honda2008crystal}
Honda,~S.; Akiba,~T.; Kato,~Y.~S.; Sawada,~Y.; Sekijima,~M.; Ishimura,~M.;
  Ooishi,~A.; Watanabe,~H.; Odahara,~T.; Harata,~K. Crystal structure of a
  ten-amino acid protein. \emph{J. Am. Chem. Soc.} \textbf{2008}, \emph{130},
  15327--15331\relax
\mciteBstWouldAddEndPuncttrue
\mciteSetBstMidEndSepPunct{\mcitedefaultmidpunct}
{\mcitedefaultendpunct}{\mcitedefaultseppunct}\relax
\EndOfBibitem
\bibitem[Buch \latin{et~al.}(2010)Buch, Harvey, Giorgino, Anderson, and
  De~Fabritiis]{buch2010high}
Buch,~I.; Harvey,~M.~J.; Giorgino,~T.; Anderson,~D.~P.; De~Fabritiis,~G.
  High-throughput all-atom molecular dynamics simulations using distributed
  computing. \emph{J. Chem. Inf. Model} \textbf{2010}, \emph{50},
  397--403\relax
\mciteBstWouldAddEndPuncttrue
\mciteSetBstMidEndSepPunct{\mcitedefaultmidpunct}
{\mcitedefaultendpunct}{\mcitedefaultseppunct}\relax
\EndOfBibitem
\bibitem[Harvey \latin{et~al.}(2009)Harvey, Giupponi, and
  Fabritiis]{harvey2009acemd}
Harvey,~M.~J.; Giupponi,~G.; Fabritiis,~G.~D. ACEMD: accelerating biomolecular
  dynamics in the microsecond time scale. \emph{J. Chem. Theory Comput.}
  \textbf{2009}, \emph{5}, 1632--1639\relax
\mciteBstWouldAddEndPuncttrue
\mciteSetBstMidEndSepPunct{\mcitedefaultmidpunct}
{\mcitedefaultendpunct}{\mcitedefaultseppunct}\relax
\EndOfBibitem
\bibitem[Piana \latin{et~al.}(2011)Piana, Lindorff-Larsen, and
  Shaw]{piana2011robust}
Piana,~S.; Lindorff-Larsen,~K.; Shaw,~D.~E. How robust are protein folding
  simulations with respect to force field parameterization? \emph{Biophys. J.}
  \textbf{2011}, \emph{100}, L47--L49\relax
\mciteBstWouldAddEndPuncttrue
\mciteSetBstMidEndSepPunct{\mcitedefaultmidpunct}
{\mcitedefaultendpunct}{\mcitedefaultseppunct}\relax
\EndOfBibitem
\bibitem[Doerr and De~Fabritiis(2014)Doerr, and De~Fabritiis]{doerr2014fly}
Doerr,~S.; De~Fabritiis,~G. On-the-fly learning and sampling of ligand binding
  by high-throughput molecular simulations. \emph{J. Chem. Theory Comput.}
  \textbf{2014}, \emph{10}, 2064--2069\relax
\mciteBstWouldAddEndPuncttrue
\mciteSetBstMidEndSepPunct{\mcitedefaultmidpunct}
{\mcitedefaultendpunct}{\mcitedefaultseppunct}\relax
\EndOfBibitem
\bibitem[Unke and Meuwly(2019)Unke, and Meuwly]{Unke_Meuwly_2019}
Unke,~O.~T.; Meuwly,~M. PhysNet: A Neural Network for Predicting Energies,
  Forces, Dipole Moments, and Partial Charges. \emph{J. Chem. Theory Comput.}
  \textbf{2019}, \emph{15}, 3678–3693\relax
\mciteBstWouldAddEndPuncttrue
\mciteSetBstMidEndSepPunct{\mcitedefaultmidpunct}
{\mcitedefaultendpunct}{\mcitedefaultseppunct}\relax
\EndOfBibitem
\bibitem[Th{\"o}lke and De~Fabritiis(2022)Th{\"o}lke, and
  De~Fabritiis]{tholke2022equivariant}
Th{\"o}lke,~P.; De~Fabritiis,~G. Equivariant transformers for neural network
  based molecular potentials. International Conference on Learning
  Representations. 2022\relax
\mciteBstWouldAddEndPuncttrue
\mciteSetBstMidEndSepPunct{\mcitedefaultmidpunct}
{\mcitedefaultendpunct}{\mcitedefaultseppunct}\relax
\EndOfBibitem
\bibitem[Sch{\"u}tt \latin{et~al.}(2019)Sch{\"u}tt, Kessel, Gastegger, Nicoli,
  Tkatchenko, and
  M{\"u}ller]{Schutt_Kessel_Gastegger_Nicoli_Tkatchenko_Muller_2019}
Sch{\"u}tt,~K.~T.; Kessel,~P.; Gastegger,~M.; Nicoli,~K.~A.; Tkatchenko,~A.;
  M{\"u}ller,~K.-R. SchNetPack: A Deep Learning Toolbox For Atomistic Systems.
  \emph{J. Chem. Theory Comput.} \textbf{2019}, \emph{15}, 448–455\relax
\mciteBstWouldAddEndPuncttrue
\mciteSetBstMidEndSepPunct{\mcitedefaultmidpunct}
{\mcitedefaultendpunct}{\mcitedefaultseppunct}\relax
\EndOfBibitem
\bibitem[{R Core Team}(2021)]{R}
{R Core Team}, R: A Language and Environment for Statistical Computing. R
  Foundation for Statistical Computing: Vienna, Austria, 2021\relax
\mciteBstWouldAddEndPuncttrue
\mciteSetBstMidEndSepPunct{\mcitedefaultmidpunct}
{\mcitedefaultendpunct}{\mcitedefaultseppunct}\relax
\EndOfBibitem
\bibitem[Wickham(2011)]{wickham2011ggplot2}
Wickham,~H. ggplot2. \emph{Wiley Interdiscip. Rev. Comput. Stat.}
  \textbf{2011}, \emph{3}, 180--185\relax
\mciteBstWouldAddEndPuncttrue
\mciteSetBstMidEndSepPunct{\mcitedefaultmidpunct}
{\mcitedefaultendpunct}{\mcitedefaultseppunct}\relax
\EndOfBibitem
\bibitem[Dowle and Srinivasan(2022)Dowle, and Srinivasan]{dowle2019package}
Dowle,~M.; Srinivasan,~A. data.table: Extension of `data.frame`. 2022; R
  package version 1.14.6\relax
\mciteBstWouldAddEndPuncttrue
\mciteSetBstMidEndSepPunct{\mcitedefaultmidpunct}
{\mcitedefaultendpunct}{\mcitedefaultseppunct}\relax
\EndOfBibitem
\bibitem[Barua \latin{et~al.}(2008)Barua, Lin, Williams, Kummler, Neidigh, and
  Andersen]{barua2008trp}
Barua,~B.; Lin,~J.~C.; Williams,~V.~D.; Kummler,~P.; Neidigh,~J.~W.;
  Andersen,~N.~H. The Trp-cage: optimizing the stability of a globular
  miniprotein. \emph{Protein Eng. Des. Sel.} \textbf{2008}, \emph{21},
  171--185\relax
\mciteBstWouldAddEndPuncttrue
\mciteSetBstMidEndSepPunct{\mcitedefaultmidpunct}
{\mcitedefaultendpunct}{\mcitedefaultseppunct}\relax
\EndOfBibitem
\end{mcitethebibliography}
\end{document}